%% file: paper303.tex
\documentclass[12pt]{article}
\usepackage{cite}
\usepackage{graphicx}
\usepackage{a4p}
\usepackage{physics}
\usepackage{l3_titlePH,ifthen,Lep}

\journalname{Eur.~Phys.~J.~C}

\date{September 23, 2005}

\preprint{2005-043}

\Lep{2}
\input{paper303.def}
\begin{document}
\begin{titlepage}
\title{ Measurement of the Mass and the Width \\
        of the W Boson at LEP }
\author{The L3 Collaboration}
\begin{abstract}
  The mass and the total decay width of the W boson are measured
  with the L3 detector at the LEP $\EE$ collider using W-boson pairs 
  produced in 0.7~fb$^{-1}$ of data collected at centre-of-mass
  energies between 161 and 209~$\GeV$.  
  Combining semi-leptonic and fully-hadronic final states, 
  the mass and the width of the W boson are determined to be
  \begin{eqnarray*}
     \MW & = &    80.270   \pm 0.046   \pm 0.031   ~\GeV \quad \mbox{and}    \\
     \GW & = & \pz 2.18\pz \pm 0.11\pz \pm 0.09\pz ~\GeV \quad ,
  \end{eqnarray*}
  where the first uncertainty is statistical and the second systematic.
\end{abstract}
%
%
%
\submitted
\end{titlepage}

\section{Introduction}

The mass, $\MW$, and the total decay width, $\GW$, are fundamental properties 
of the W boson.
Their measurement, initially performed at the S$\mathrm{p \bar p}$S hadron
collider~\cite{spps-mw-gw}, provides important information about the
Standard Model of electroweak interactions~\cite{standard-model}. 
Together with other electroweak parameters such as the Z-boson mass,
the effective weak mixing angle
and the measurement of the top-quark mass~\cite{tevatron-mtop},
the precise determination of $\MW$ allows a thorough test of the 
Standard Model at the quantum loop level
as well as constraining the mass of the Higgs boson~\cite{ewwg}.

In ${\rm e^+e^-}$ collisions, W bosons are produced singly or in pairs.
At centre-of-mass energies, $\sqrt{s}$, exceeding $2\MW$, 
W-boson pair production, ${\rm e^+e^- \to W^+W^-}$, dominates. 
The pair-production cross section at threshold is sensitive to $\MW$.
Therefore, at LEP $\MW$ was first derived from 
cross section measurements~\cite{l3-threshold-mw,lep-threshold-mw}.
At centre-of-mass energies well above production threshold, 
W bosons are directly reconstructed and the effective mass 
of the decay products is used to determine 
$\MW$~\cite{l3-direct-mw,lep-direct-mw}.
The mass distribution of the W bosons is analysed and $\MW$ and $\GW$
are determined by comparing samples of Monte Carlo events to data.  
A reweighting procedure is applied to obtain Monte Carlo samples 
corresponding to different values of $\MW$ and $\GW$.  

In the following, $\MW$ and $\GW$ are defined such that the
denominator of the W-boson propagator, $(m^2-\MW^2)+im^2\GW/\MW$,
models the mass-dependent width of the W boson.
The analysis presented here is based on a data sample collected with the 
L3 detector~\cite{l3-detector} at $\sqrt{s} = 189-209~\GeV$, 
corresponding to a total integrated luminosity of 629~pb$^{-1}$.
These results are combined with previous L3 measurements at lower 
centre-of-mass energies~\cite{l3-threshold-mw,l3-direct-mw} yielding 
final results on $\MW$ and $\GW$ based on the complete data sample of 
706~pb$^{-1}$ collected by the L3 experiment at  $\sqrt{s} = 161-209~\GeV$.
Other measurements of $\MW$ at LEP and the TEVATRON are described in
References~\citen{lep-direct-mw} and~\citen{tevatron-mw}, respectively.
The indirect determination of $\MW$ from electroweak precision data
is presented in Reference~\citen{ewwg}.

\section{Data sample}

W bosons decay into hadrons, mostly through ${\rm W^- \to
\bar{u}d~or~\bar{c}s}$, or leptons, ${\rm W^- \to
\ell^-\bar\nu_\ell}$, where $\ell$ denotes an electron, muon or tau
lepton.  Charge-conjugate states are understood to be included
throughout this article.  In the following, these final states are
denoted as $\QQ$ and $\LN$, or, in general, $\FF$, for both $\rm W^+$
and $\rm W^-$ decays.  W-boson pair production yields three classes of
events: the fully-leptonic, $\LNLN$, the semi-leptonic, $\QQLN$, and
the fully-hadronic, $\QQQQ$, final states.  Due to the presence of
more than one neutrino in the $\LNLN$ final state, the effective
masses of the W bosons cannot be directly reconstructed from their
decay products and this decay channel is not further considered here.
Visible final-state fermions are reconstructed in each event.
Electrons and muons from W-boson or $\tau$-lepton decays are measured
in the calorimeters and in the tracking system.  Hadronically-decaying
$\tau$-leptons are identified as narrow, low-multiplicity, jets.  Jets
originating from quarks are reconstructed by combining information
from calorimetric clusters and associated tracks into jets using the
DURHAM algorithm~\cite{DURHAM}.

The data analysed correspond to seven average values of $\sqrt{s}$, 
listed in Table~\ref{tab:event-sample}.
The selection of W-boson pair-production events is described in 
Reference~\citen{l3-291-ww-xsection}.
The selection of the $\QQEN$ and $\QQMN$ final states requires
an identified high-energy electron or muon, respectively.
The $\QQTN$ final state is characterised by a low-energy isolated
electron or muon or by the reconstruction of a narrow jet.
For all semi-leptonic final states, missing momentum due to the 
neutrino is required and the jet-jet mass has to be compatible with $\MW$.
The selection of the $\QQQQ$ final state requires events with high 
multiplicity, small missing energy and a four-jet topology.
To reject quark-pair production with additional jets originating from 
radiated gluons an artificial neural network is trained using 
discriminating variables such as the jet energies, broadenings and angles,
the event spherocity, the jet multiplicity and the DURHAM jet-resolution 
parameter for which the event topology changes from three to four jets, 
$y_{34}$.
Only events with high neural-network output are retained for further analysis.
The numbers of selected W-boson pairs are detailed in 
Table~\ref{tab:event-sample}.

\section{Monte Carlo simulation}

The KANDY~\cite{KANDY} Monte Carlo generator is used to model
four-fermion production, including both W-boson production 
and background processes.
The RACOONWW~\cite{RACOON} program is used as a cross check and
to estimate possible systematic uncertainties due to the modelling
of photon radiation.
Additional background contribution from fermion-pair production, 
dominated by the $\EE\to\QQ$ process, is simulated using the 
KK2F~\cite{KK2F} event generator. 
Monte Carlo events are generated at the seven average $\sqrt{s}$ values
listed in Table~\ref{tab:event-sample}.
Effects from the spread of centre-of-mass energies within the
individual energy points are found to be negligible.
The expected number of background events is listed in 
Table~\ref{tab:event-sample}.

The hadronisation process is modelled with the PYTHIA~\cite{PYTHIA}
program, while the HERWIG~\cite{HERWIG} and ARIADNE~\cite{ARIADNE}
programs are used to assess systematic uncertainties.  These Monte
Carlo programs are tuned to describe hadronic Z-boson decays recorded
at the Z resonance~\cite{l3-290-qcd-report}.  In the case of W-boson
pair production, a dedicated parameter set, tuned only on Z-boson
decays into light-quarks (u,d,c,s) is used. 

Bose-Einstein correlations (BEC)~\cite{FSI-BE} in W-boson decays are simulated
using the BE32 model~\cite{MC-BE} implemented in PYTHIA.
Only BEC between hadrons originating from the same W boson are taken
into account, as suggested by our measurements~\cite{l3-257-BE}. 
Colour-Reconnection (CR) effects~\cite{FSI-CR} in the $\QQQQ$ final state
would alter the colour flow between the W bosons. 
In accordance with our measurement~\cite{l3-266-CR}, these are not 
implemented in the Monte Carlo simulation.
However, both the effect of BEC between hadrons originating from different 
W decays and that of CR between W bosons are considered as possible systematic 
uncertainties.

The response of the L3 detector is modelled with the GEANT~\cite{GEANT}
program which includes the effects of energy loss, 
multiple scattering and showering in the detector material. 
Hadronic showers are simulated with the GHEISHA~\cite{GHEISHA} program. 
Time-dependent detector efficiencies, as monitored during data taking, 
are included in the simulation.

\section{Event reconstruction}
\label{sec:reco-kf}

In the $\QQEN$, $\QQMN$ and $\QQQQ$ channels a kinematic fit is applied to 
improve the resolution of the measured energies, $E_f$, momenta, $p_f$, 
polar, $\theta_f$, and azimuthal, $\phi_f$, angles of the visible fermions.
Four-momentum conservation and other constraints, as detailed below,
are imposed.
The measured quantities are varied within their resolution to 
satisfy these constraints.
The resolution of each individually-measured object depends on details 
of the reconstruction, such as the detector region or the energy scale. 
The average resolutions of $E_f$, $\theta_f$, and $\phi_f$ for electrons, 
muons and hadronic jets, as determined by Monte Carlo simulation,
are given in Table~\ref{tab:object-resolution}.
These values agree with the resolutions derived from calibration data 
collected at the Z resonance within the statistical accuracy of the test.
 
In all events, four-momentum conservation is required which, in the case of the 
$\QQEN$ and $\QQMN$ final states, determines the momentum and the direction of 
the neutrino.  
For hadronic jets, the velocity $\beta_f=p_f/E_f$ is fixed to its 
measured value, as many systematic effects cancel in this ratio. 
When imposing energy conservation, the $\sqrt{s}$ value determined for 
each event by the LEP Energy Working Group~\cite{lep-energy} is used. 
Events collected during the manipulation of the LEP beams,
for which no precise calibration of the LEP energy is available,
are excluded from the analysis.
Energy conservation results in a one-constraint (1C) kinematic fit for 
$\QQEN$ and $\QQMN$ final states and a four-constraint (4C) kinematic fit 
for the $\QQQQ$ channel.
In general, fermion angles are better measured than energies and momenta. 
Therefore, the kinematic fit improves more the determination of the latter. 
The improvement in the resolution of the average value of the two
reconstructed W-boson masses is shown in Table~\ref{tab:mass-resolution}.

In the 1C and 4C fits, the masses of the two W bosons are determined
separately.  The mass resolution is further improved by the additional
constraint of requiring these masses to be equal within the width of the W boson, fixed
as $2.1~\GeV$. This numerical value does not bias the resulting fits.  This
procedure results in a two-constraint (2C) fit of $\QQEN$ and $\QQMN$ events and
a five-constraint (5C) fit of $\QQQQ$ events.  For $\QQEN$ and $\QQMN$
events, both the mass of the hadronically-decaying W boson obtained in
the 1C fit, $\Minvone$, and the average mass obtained in the 2C fit,
$\Minvtwo$, are used in the mass extraction, which is described in the
following section.  Similarly, in the $\QQQQ$ channel, the average
masses of the 4C fit, $\Minvfour$, and the 5C fit, $\Minvfive$, are
used.

The $\QQTN$ final state contains at least two neutrinos and
only the W boson decaying into hadrons is used in the mass reconstruction.  
The energies of the two hadronic jets are rescaled by a common factor such 
that the sum of their energies equals $\sqrt{s}/2$, effectively imposing 
an equal-mass constraint on the two W bosons.  
Use of the mass of the hadronic system after this rescaling, $m_\mathrm{resc}$, 
improves the resolution of the W-boson mass reconstruction by more than 
a factor of two.

\begin{figure}
  \includegraphics[width=0.49\textwidth]{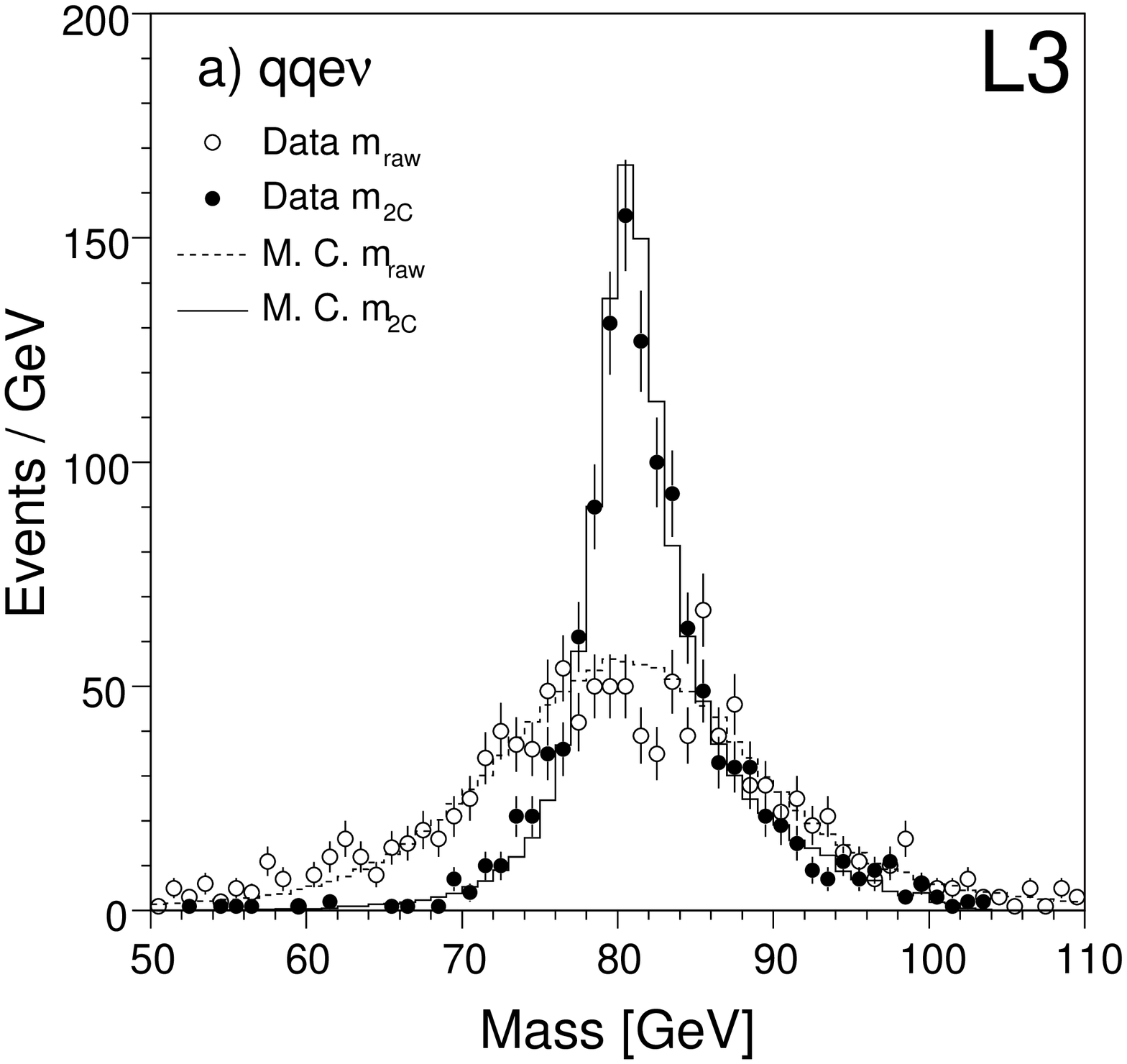}
  \hfill
  \includegraphics[width=0.49\textwidth]{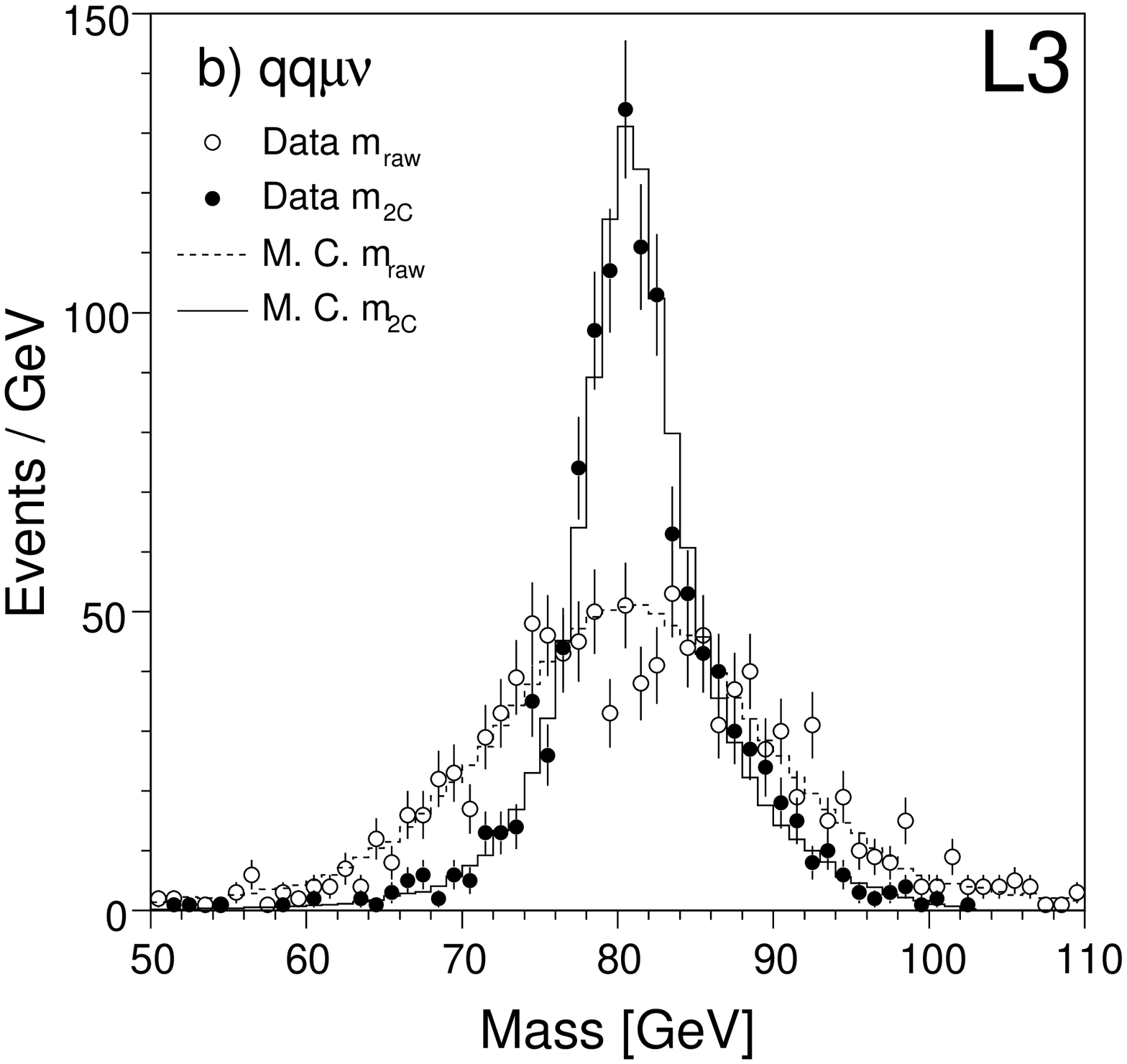}

  \includegraphics[width=0.49\textwidth]{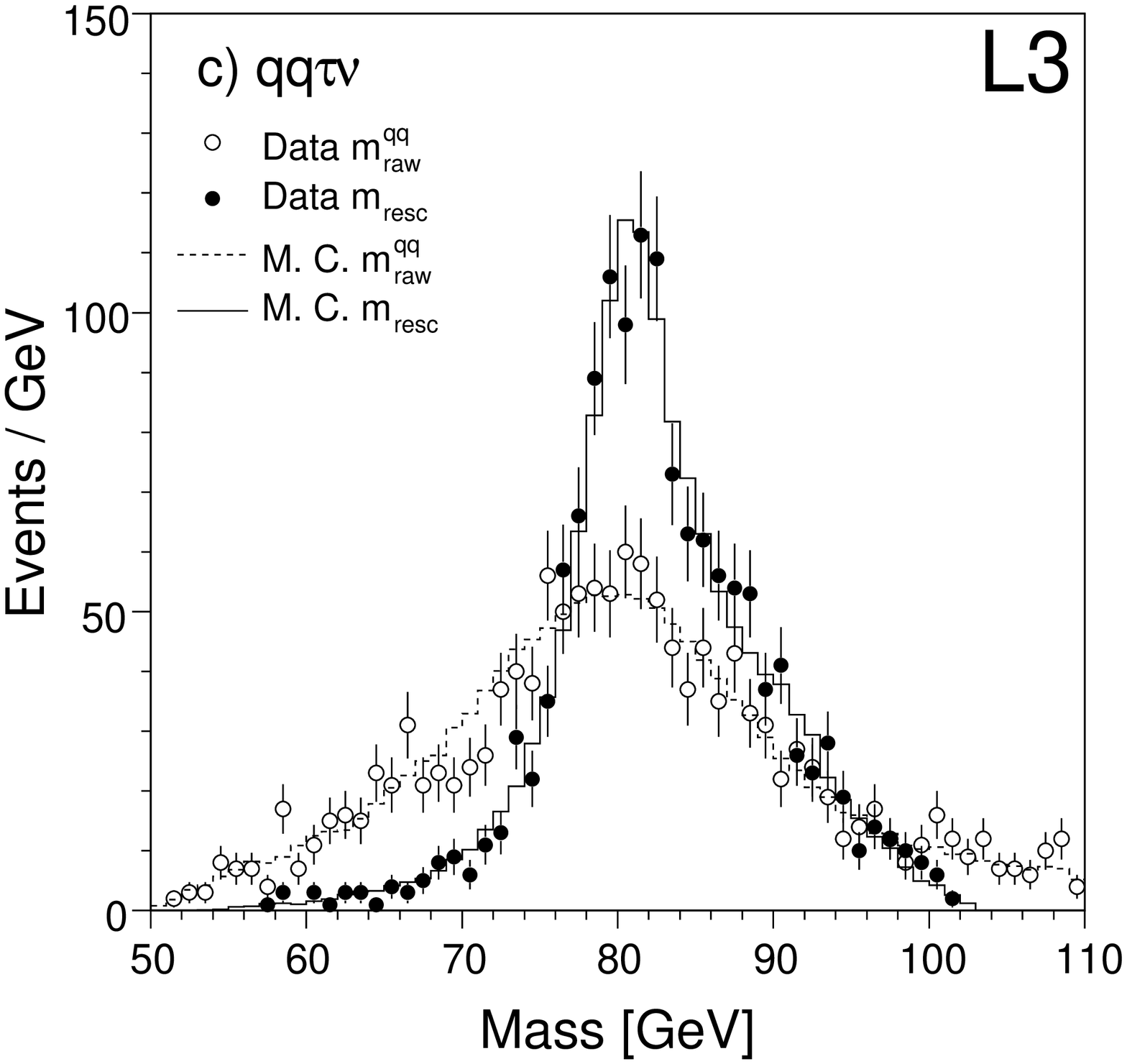}
  \hfill
  \includegraphics[width=0.49\textwidth]{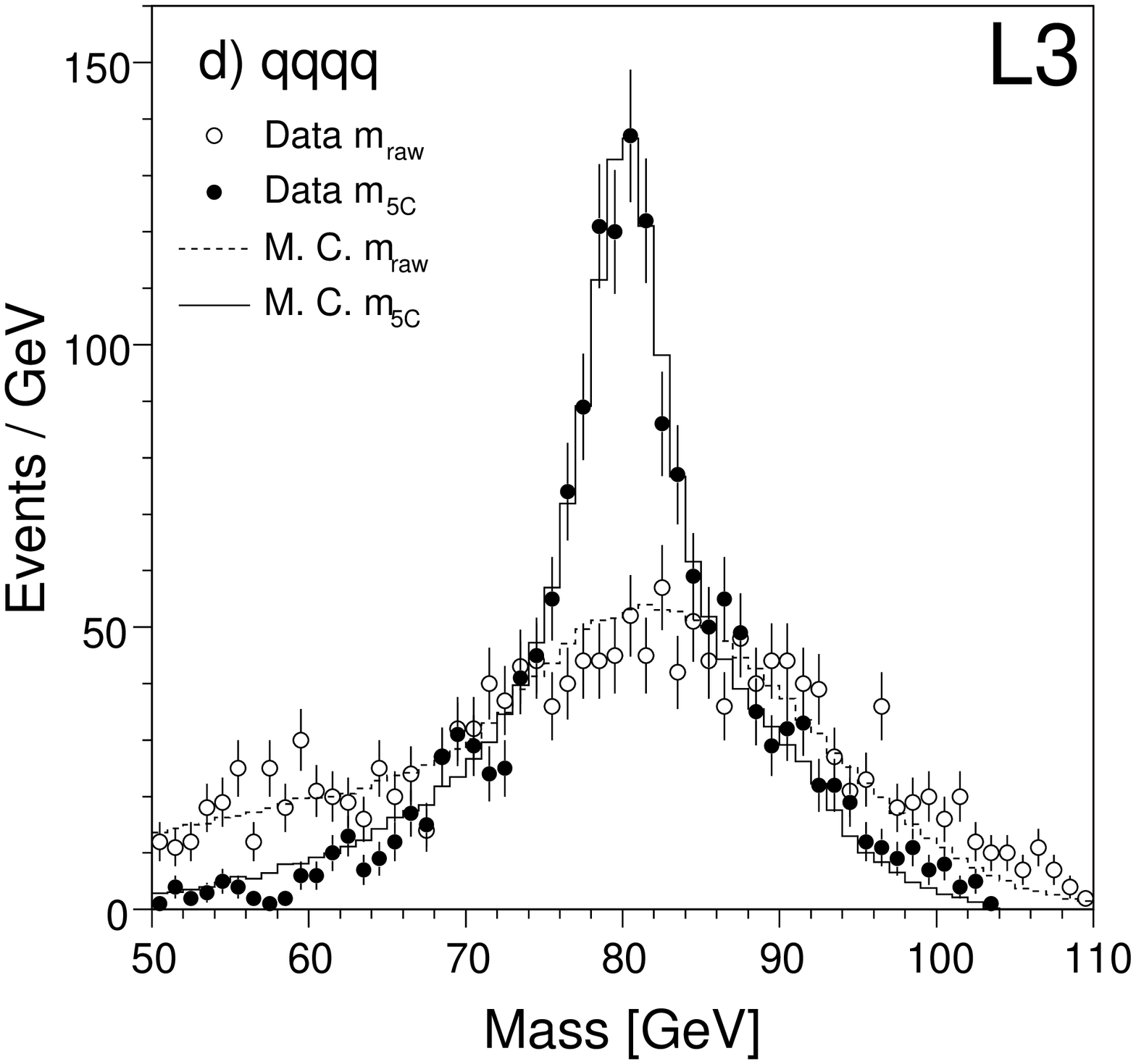}
  \caption{
  Improvement of mass resolutions due to kinematic constraints for
  a) $\QQEN$, b) $\QQMN$, c) $\QQTN$ and d) $\QQQQ$ events.
  The open circles represent the raw mass spectra and 
  the full points the spectra obtained after applying the kinematic fit or 
  the jet-energy rescaling.
  Monte Carlo predictions are also shown. In a), b) and d) $m_{\rm raw}$
  is the average
  of the two raw masses while in c) $m^{\rm qq}_{\rm raw}$ is the raw mass of the
  hadronic system.
  }
\label{fig:mass-resolution}
\end{figure}

The improvement of the mass resolution due to the kinematic fit 
is shown in Figure~\ref{fig:mass-resolution}.
The average mass resolutions before and after the kinematic fits 
or the energy rescaling is summarised in Table~\ref{tab:mass-resolution} 
for all final states.
Only the better-measured quantities $\Minvtwo$ and $\Minvfive$ are used to
determine $\GW$.

W-boson pair production is frequently accompanied by photon radiation.
Photons near to a final-state fermion are mainly due to 
final-state radiation (FSR).
In $\QQEN$ events, photons close to the electron are automatically included 
into the measurement of the electromagnetic cluster.
In $\QQMN$ events, the cluster closest to the muon direction 
is assumed to originate from the ionisation energy loss of the muon 
in the calorimeters and is taken out of the event.
In $\QQTN$ events, the photon clusters are combined in the tau jet 
by the jet-reconstruction cone algorithm. 
Hard photons with energies greater than $5~\GeV$ and outside a cone of 
$5^\circ$ half-opening angle around the lepton are detected
in 5\% of the $\QQEN$ events and in 2.5\% of the $\QQMN$ events.
They are taken into account by the kinematic fit, but not incorporated in the 
mass reconstruction, as they are mainly due to initial-state radiation (ISR).
In all other cases, the detected photons are assigned to the jets during
the clustering process.
For photons emitted along the beam direction, and therefore undetected,
the analysis relies on the Monte Carlo simulation.

Systematic uncertainties arise in the $\QQQQ$ channel due to potential 
effects of CR between the jets from different W bosons.
To reduce these effects, clusters with an energy below $2~\GeV$ are removed 
from the original jets obtained by the jet clustering process,
as discussed in Section~\ref{sec:syst-fsi}.
The jet energies and momenta are re-scaled with an equal scale factor
in order to obtain the original jet energy.
Only the jet directions and the jet masses are affected by this procedure
and the energy resolution of the jets is preserved as illustrated in
Table~\ref{tab:object-resolution}.
On the other hand, the angular resolution of the jets is worsened leading to
a degradation of the W-boson mass resolution by about 20\%.
The resulting increase of the statistical uncertainty on $\MW$
is overcompensated by a reduction of the systematic uncertainty,
leading to a lower total uncertainty, as discussed in Section~\ref{sec:syst-fsi}.

The mass resolution in the fully-hadronic final state is
improved by taking into account gluon radiation from quarks. 
The DURHAM jet-resolution parameter for which the event topology changes
from four to five jets, $y_{45}$, is used to separate events 
with and without gluon radiation. 
Those with $\log{y_{45}} > -6.2$ are treated as four-jet events and the 
remaining as five-jet events. 

The four or five jets must be associated with the two W bosons. 
In the case of four jets, all three combinations are considered. 
Five jets can be paired in ten different ways. 
Monte Carlo studies show that in five-jet events only some combinations 
have a high probability to be correct.
These are the ones in which the W boson that decays without hard-gluon 
radiation is formed by the highest-energy jet and any other jet or by 
the second highest-energy jet and any other jet except the lowest-energy jet.
Only these six combinations are considered.
The three pairings with the highest kinematic-fit probability are retained.
They are ordered by their fit probability and treated as separate samples. 
Pairings where the fit did not converge are rejected.
This criterion removes 5\% of the events.

The four-jet and five-jet samples, of about equal size, are treated in 
separate mass fits since their mass resolutions are different by about 
30\% as shown in Table~\ref{tab:mass-resolution}.
Due to the overall improvement in mass resolution, the statistical uncertainty of 
$\MW$, as determined in the fully-hadronic channel, is reduced by 6\%.

\begin{figure}
  \includegraphics[width=0.49\textwidth]{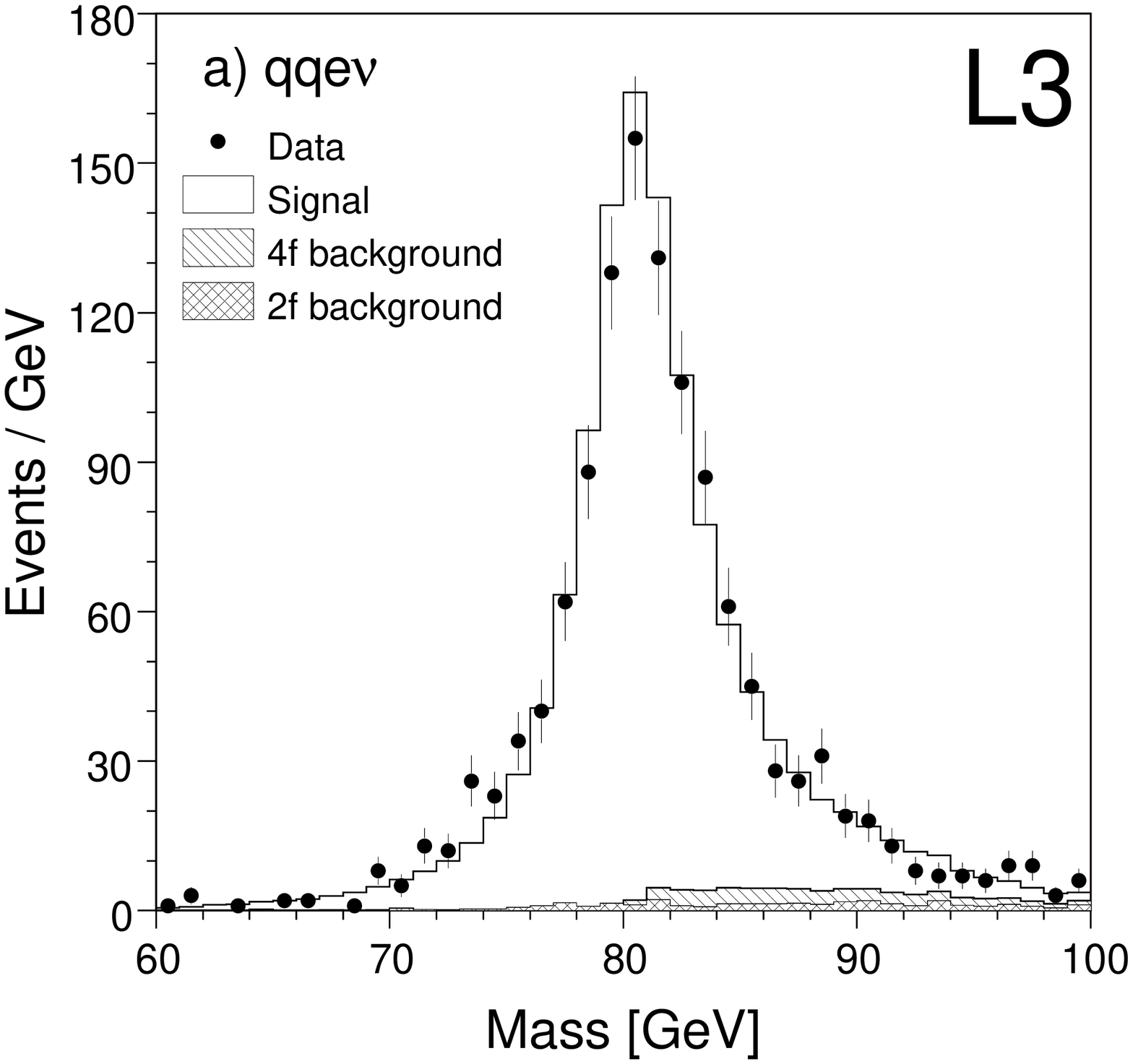}
  \hfill
  \includegraphics[width=0.49\textwidth]{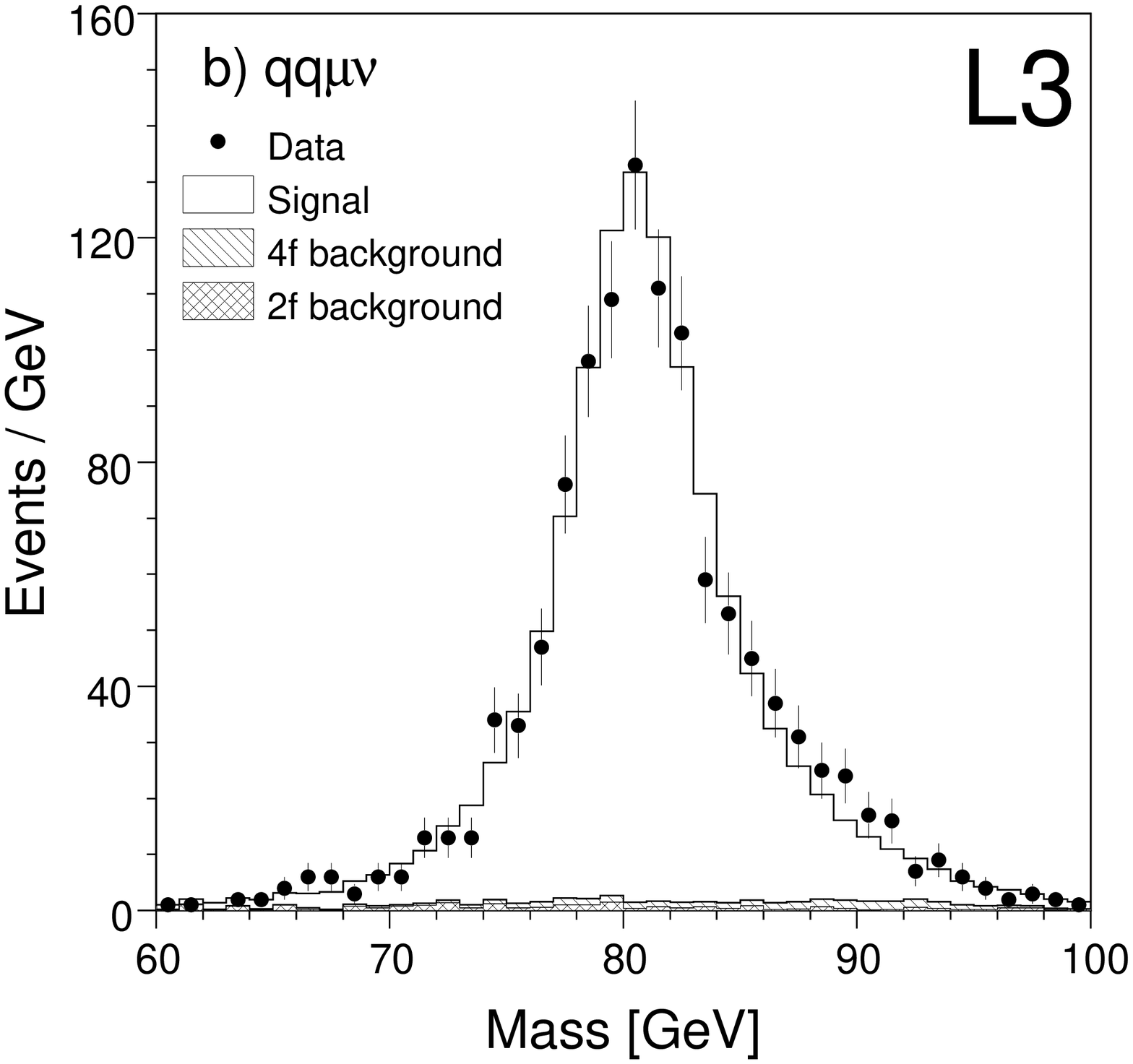}

  \includegraphics[width=0.49\textwidth]{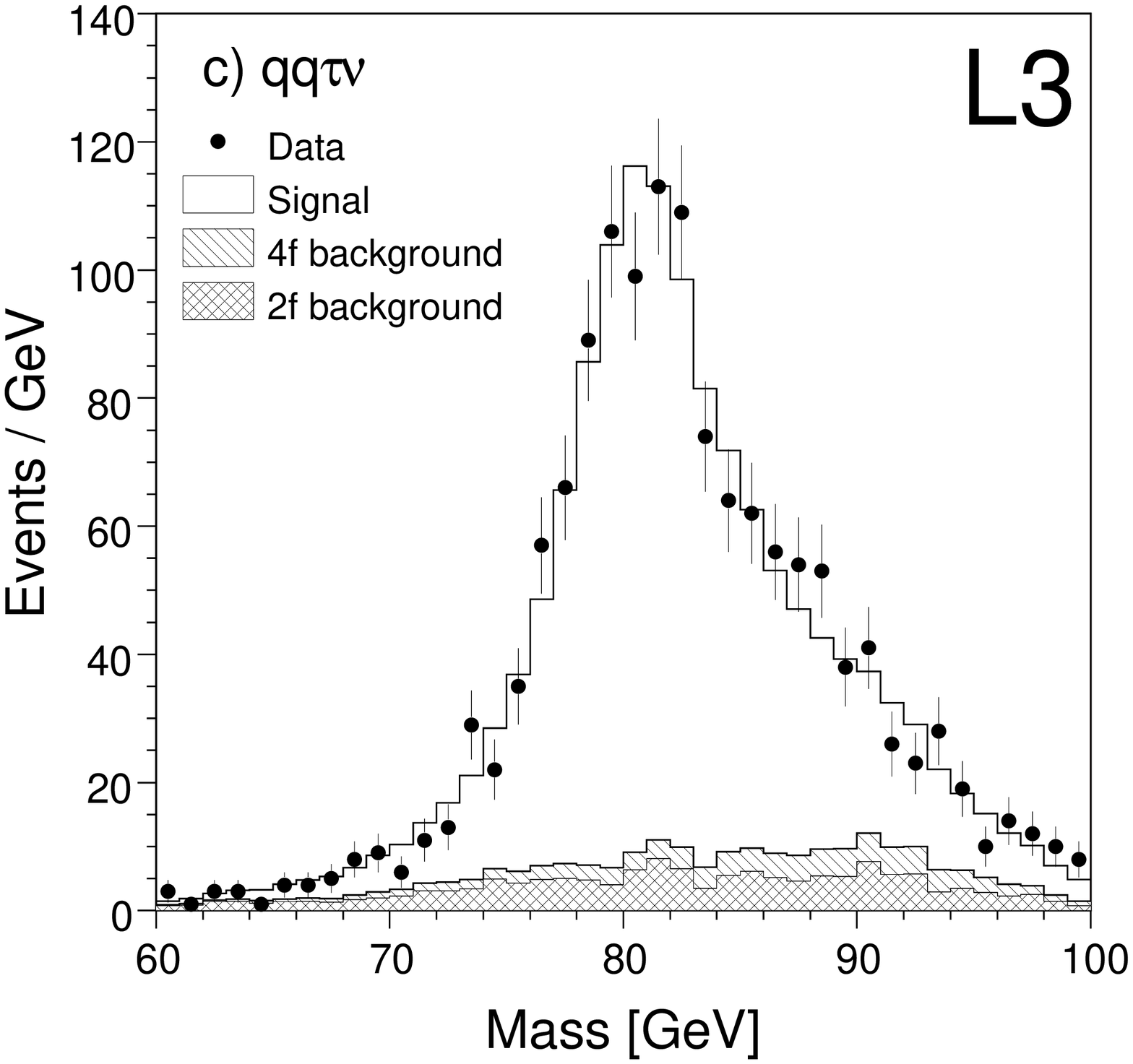}
  \hfill
  \includegraphics[width=0.49\textwidth]{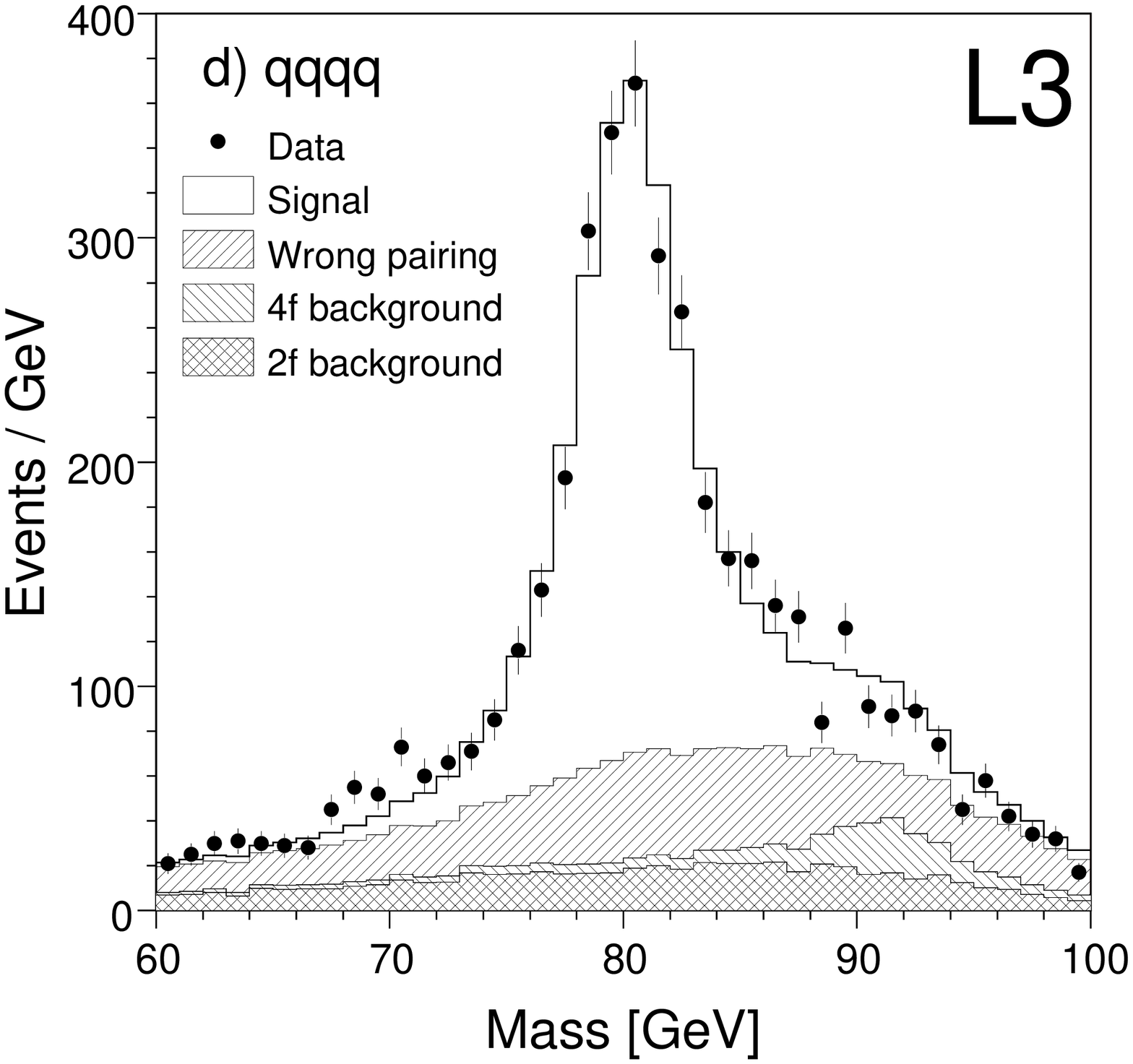}
  \caption{ 
    Distributions of reconstructed W-boson masses after applying the 
    kinematic fit using the equal-mass constraint for the
    a) $\QQEN$, b) $\QQMN$ and c) $\QQTN$ channels and d) the best
    pairing for the $\QQQQ$ channel.
    The signal Monte Carlo events are reweighted according to the fitted
    value of $\MW$.
  }
\label{fig:mw-minv-4}
\end{figure}

\begin{figure}
  \centerline{\includegraphics[width=0.95\textwidth]{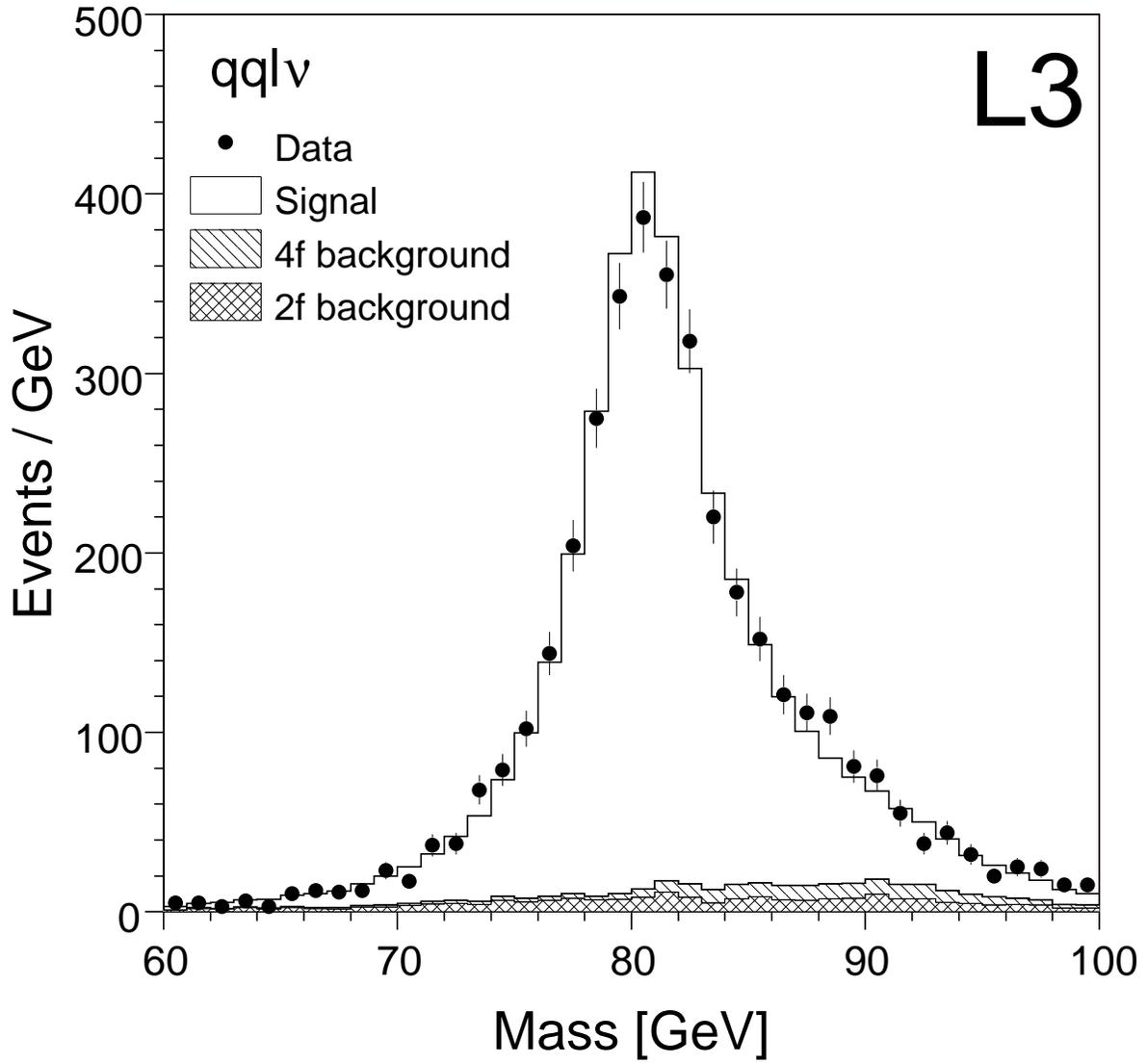}}
  \caption{
    Distribution of reconstructed W-boson masses after applying the 
    kinematic fit using the equal-mass constraint for 
    semi-leptonic final states.
    The signal Monte Carlo events are reweighted according to the fitted
    value of $\MW$.
  }
\label{fig:mw-minv-qqln}
\end{figure}

\begin{figure}
  \centerline{\includegraphics[width=0.95\textwidth]{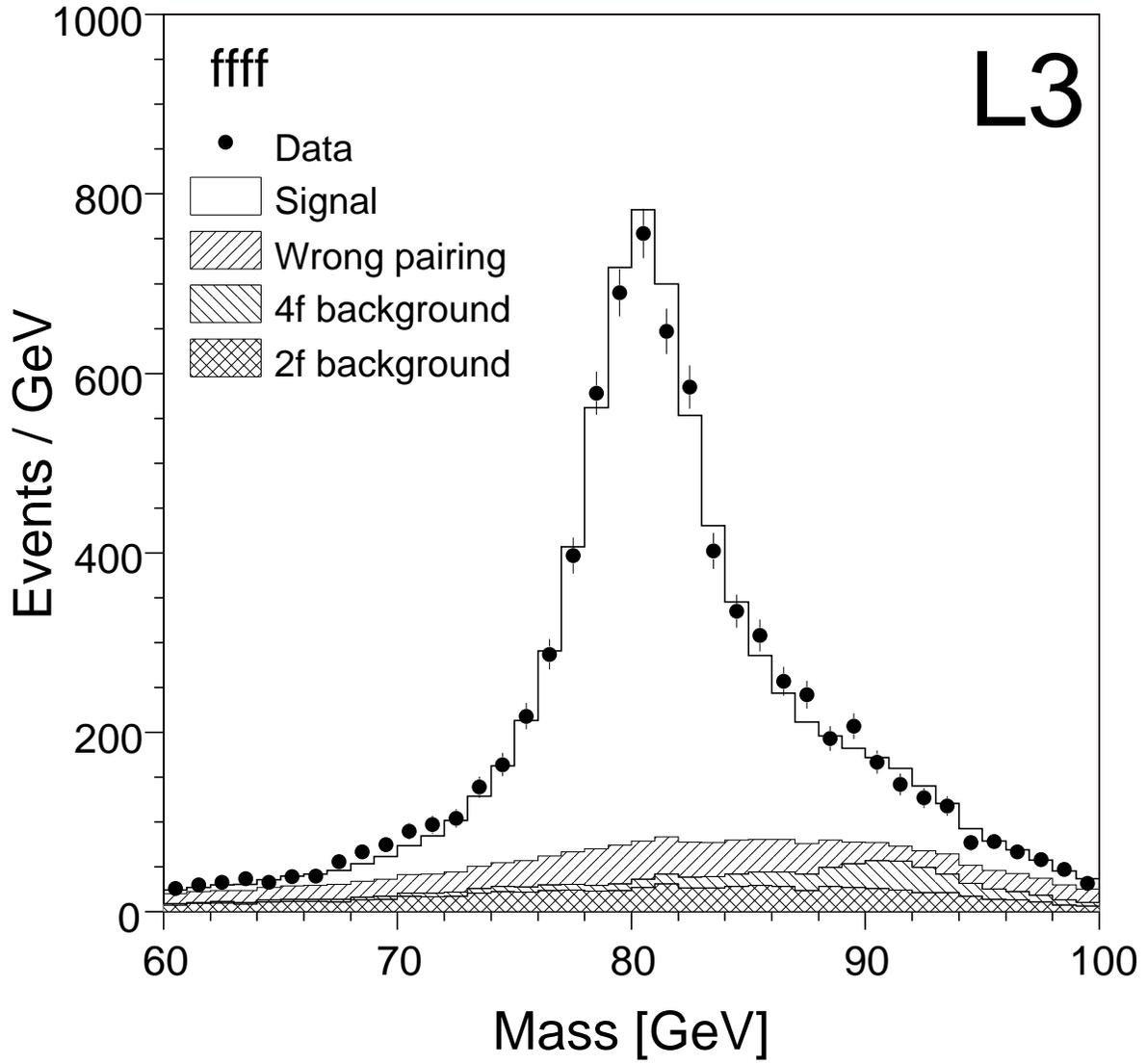}}
  \caption{
    Distribution of reconstructed W-boson masses after applying the 
    kinematic fit using the equal-mass constraint for all W pairs.
    The signal Monte Carlo events are reweighted according to the fitted
    value of $\MW$.
  }
\label{fig:mw-minv-ffff}
\end{figure}

The mass spectra after the kinematic fit for the better-measured mass 
variable $m_1$ are shown in Figure~\ref{fig:mw-minv-4} for the 
semi-leptonic final states and the best pairing in the 
fully-hadronic final state.
Figure~\ref{fig:mw-minv-qqln} presents the sum of the
semi-leptonic distributions, 
while Figure~\ref{fig:mw-minv-ffff} shows the sum of all
four distributions.

\section{\boldmath Extraction of $\MW$ and $\GW$}
\label{sec:fit}

A maximum-likelihood method is used to extract $\MW$ and $\GW$ from
the reconstructed masses of each event. 
The extraction of $\MW$ and $\GW$ is done separately for each 
of the four final states, $\QQEN$, $\QQMN$, $\QQTN$ and $\QQQQ$,
and the seven average values of $\sqrt{s}$. 
For each of these 28 event samples a likelihood function, $L(\MWfit,\GWfit)$, 
is constructed from the product of the individual likelihoods.
These are evaluated for each mass reconstruction, $i$, performed for a given 
semi-leptonic event or a given pairing of the four- and five-jet samples of 
the fully-hadronic final state.
Correlations between the reconstructed masses from different pairings
are found to be negligible.
The individual likelihoods are calculated from the normalised differential 
cross sections in terms of the reconstructed masses, $m_1$ and $m_2$,
\begin{eqnarray}
L(\MWfit,\GWfit) & = & 
\prod_{\textstyle i} \; 
\frac{\displaystyle f(\MWfit,\GWfit) \, 
      \left(\frac{\d^2\sigma(\MWfit,\GWfit)}{\d m_1 \; \d m_2}\right)_{\textstyle i} 
      + \left(\frac{\d^2\sigma_{\BG}}{\d m_1 \; \d m_2}\right)_{\textstyle i}
      }
      {\Bigg. f(\MWfit,\GWfit)\,\sigma(\MWfit,\GWfit)+\sigma_{\BG}} \; ,
\end{eqnarray}
where $\sigma$ and $\sigma_{\BG}$ are the accepted signal and
background cross sections of the corresponding final state. 
As summarised in Table~\ref{tab:mass-resolution}, the masses are chosen as
$m_1 = \Minvtwo$ and $m_2 = \Minvone$ for $\QQMN$ and $\QQEN$ final states and
$m_1 = \Minvfive$ and $m_2 = \Minvfour$  for fully-hadronic events.
For $\QQTN$ events the doubly-differential cross section
is reduced to a singly-differential one and only the rescaled mass of the 
hadronic system is used, $m_1 = m_\mathrm{resc}$.
The normalisation factor $f(\MWfit,\GWfit)$ is calculated such that the sum of
the accepted background and the reweighted signal cross section reproduces
the measured cross section. 
This procedure determines $\MW$ and $\GW$ solely from the shapes of the 
mass distributions.
In the fits to determine $\MW$, the Standard Model relation 
$\GW = 3\GF\MW^3(1+2\aqcd/3\pi)/(2\sqrt{2}\pi)$ is imposed~\cite{lep2-yr-ww}.
When $\GW$ is extracted, $\MW$ and $\GW$ are treated as independent quantities
and the doubly-differential cross section is reduced to a single one, 
since only the better-measured quantity $m_1$ is used for the determination
of $\GW$.

The total and differential cross sections of signal and background accepted
by the event selection are determined using Monte Carlo simulations. 
Except for single-W production, the background cross sections are 
independent of $\MW$ and $\GW$. 
The signal Monte Carlo simulation, which is originally generated using a particular 
value of the W-boson mass, $\MWgen$, and width, $\GWgen$, is modified
in a reweighting procedure to represent a different W-boson mass, 
$\MWfit$, and width, $\GWfit$.
Each signal Monte Carlo event, $j$, is given a new weight, $R_j$, defined by
the ratio
\begin{eqnarray}
R_j(\MWfit,\GWfit,\MWgen,\GWgen) & = & 
\frac
{\left|{\cal M}(p_j^1,p_j^2,p_j^3,p_j^4,k_j^{\gamma},\MWfit,\GWfit)\right|^2}
{\left|{\cal M}(p_j^1,p_j^2,p_j^3,p_j^4,k_j^{\gamma},\MWgen,\GWgen)\right|^2}
\, ,
\end{eqnarray}
where ${\cal M}$ is the matrix element of the four-fermion final state
under consideration.  The matrix elements are calculated for the
generated four-vectors of the four fermions, $p_j^{n=1...4}$ using the
program EXCALIBUR~\cite{EXCALIBUR}.  Since this program is based on
four-fermion final states without additional photons, the momentum sum
of any ISR photons present in the Monte Carlo events, $k_j^{\gamma}$,
is taken into account by boosting the four fermions into the rest
frame of the event after the ISR photon emission.  Photons not emitted
in the initial state are recombined with the closest final-state
fermion.  It was verified that this method is equivalent to using the
KANDY program, which simulates photon radiation in the event
generation.  As a cross check of the matrix element reweighting the
event weights are evaluated from a Breit-Wigner function.  Consistent
results are observed.

The total accepted signal cross section for a given set of parameters,
$\MWfit$ and $\GWfit$, is
\begin{eqnarray}
  \sigma(\MWfit,\GWfit) & = & 
  \frac{\sigma^\gen}{N^\gen} \;
  \sum_j R_j(\MWfit,\GWfit,\MWgen,\GWgen) \; ,
\end{eqnarray}
where $\sigma^{\gen}$ denotes the cross section corresponding to
the total Monte Carlo sample containing $N^{\gen}$ events. 
The sum extends over all Monte Carlo events, $j$, accepted by the event
selection.
The total background cross section is
\begin{eqnarray}
  \sigma_\BG & = & \sum_l
  \frac{\sigma^\gen_{\BG,l}}{N^\gen_{\BG,l}} \, N^\mathrm{sel}_{\BG,l} \; ,
\end{eqnarray}
where, for each background process $l$ with generated cross section
$\sigma^\gen_{\BG,l}$, $N^\gen_{\BG,l}$ and $N^\mathrm{sel}_{\BG,l}$
are the numbers of generated and accepted Monte Carlo events, respectively.

To determine the accepted differential cross section for a given 
data event, $i$, the box method~\cite{BOXMETHOD} is applied. 
When combined with the reweighting procedure, this method takes into account
detector and selection effects, efficiencies and purities which depend on
$\MW$ and $\GW$ and correlations between the input masses $m_1$ and $m_2$.
The accepted differential cross section is determined by averaging 
signal Monte Carlo events inside a two-dimensional mass domain, $\Omega_i$, 
centred around each data event. 
To take the different resolutions of $m_1$ and $m_2$ into account,
these masses are rescaled by their resolutions $\sigma_1$ and $\sigma_2$
whose averages are given in Table~\ref{tab:mass-resolution}.
The size of each domain is limited by requiring a sufficient number of 
Monte Carlo events in the domain.
In the rescaled parameter space, the distance, $d_{ij}$, of each Monte Carlo 
event, $j$, from the given data event, $i$, with reconstructed masses 
$(m_1)_i$ and $(m_2)_i$ is calculated from
\begin{equation}
d_{ij} = 
\sqrt{\left(\frac{(m_1)_j-(m_1)_i}{\sigma_1}\right)^2
     +\left(\frac{(m_2)_j-(m_2)_i}{\sigma_2}\right)^2} \; ,
\end{equation}
and the 400 closest Monte Carlo events are retained.

The most distant Monte Carlo event, $j_\mathrm{max}$, determines
the mass intervals around the data event, 
$(\delta m_1)_i = |(m_1)_i-(m_1)_{j_\mathrm{max}}|$ and 
$(\delta m_2)_i = |(m_2)_i-(m_2)_{j_\mathrm{max}}|$, which vary between 
$200~\MeV$ and $600~\MeV$.
After summing the weights $R_j$ of all Monte Carlo events associated 
to the mass domain $\Omega_i$ around the considered data event,
the differential cross section of the signal processes is given by
\begin{eqnarray}
\left(\frac{\d^2\sigma(\MWfit,\GWfit)}{\d m_1 \; \d m_2} \right)_{\textstyle i}
& = &
\frac{1}{\pi (\delta m_1)_i (\delta m_2)_i} \,
\frac{\sigma^\gen}{N^\gen} \,
\sum_{j \in \Omega_i} R_j(\MWfit,\GWfit,\MWgen,\GWgen) \; .
\end{eqnarray}

For the background Monte Carlo simulation, the same domain size as for the signal
is chosen and the differential distribution of the background is determined 
from the number of selected background Monte Carlo events, 
$(N^\mathrm{sel}_\BG)_i$, associated with a given data event:
\begin{eqnarray}
\left(\frac{\d^2\sigma_\BG}{\d m_1 \; \d m_2} \right)_{\textstyle i}
& = &
\frac{1}{\pi (\delta m_1)_i (\delta m_2)_i} \,
\frac{\sigma^\gen_\BG}{N^\gen_\BG} \, 
(N^\mathrm{sel}_\BG)_i \; .
\end{eqnarray}

One-dimensional boxes in the $m_1$ space are constructed for the 
determination of $\GW$. 
The size of each bin is defined by requiring at least 200, 
but not more than 1000, Monte Carlo events.  
The bin size is at most $\pm250~\MeV$ around $(m_1)_i$ and decreases 
to about $\pm30~\MeV$ around the peak of the mass spectrum. 
For the background Monte Carlo simulation, the bin size is chosen as $\pm1~\GeV$ 
around $(m_1)_i$.

\section{Systematic uncertainties}
\label{sec:syst-err}

The systematic uncertainties on $\MW$ and $\GW$ are summarised in 
Tables~\ref{tab:mw-syst} and~\ref{tab:gw-syst}, respectively. 
They arise from various sources correlated or un-correlated between the 
final states and between the various $\sqrt{s}$ values.
The different sources of the systematic uncertainty are detailed
in the following subsections and their correlations are discussed in
Section~\ref{sec:results}.

Systematic uncertainties are assessed by determining $\Delta\MW$ and 
$\Delta\GW$ which are defined as the changes of the $\MW$ and $\GW$ results
if alternative detector calibrations, Monte Carlo simulations or 
reconstruction procedures are used.
Two methods are used for the evaluation of $\Delta\MW$.
In the cases where the effect of an alternative Monte Carlo simulation
is studied, the usual mass fit is used, but the data events are replaced 
by a high-statistics sample from the alternative simulation.
The fit result, $\MWfit$, is compared to the nominal
W-boson mass common to both Monte Carlo samples, $\MWgen$, deriving
$\Delta\MW = \MWgen-\MWfit$.
A similar procedure is used to derive $\Delta\GW$.
In the cases where the agreement between data and simulation is analysed,
the shift of the reconstructed mass is calculated for each data and 
Monte Carlo event.
The average mass shifts of the data and Monte Carlo distributions
are compared to determine $\Delta\MW$.

\subsection{\boldmath Calibration of $\sqrt{s}$}

The value of $\sqrt{s}$ is used as a constraint in the kinematic fit.
A variation of $\sqrt{s}$ would imply a shift of the reconstructed masses. 
The relative uncertainty on $\MW$ is the same as that on $\sqrt{s}$, 
while $\GW$ is less affected.  
This is verified by comparing simulated event samples in which the 
$\sqrt{s}$ value used in the kinematic fit is systematically changed.
The dependences of $\MW$ and $\GW$ on $\sqrt{s}$ are taken 
into account using the LEP energy determined for the exact time each 
W-boson pair was recorded.
The LEP beam energy is known with an accuracy between $10$ and 
$20~\MeV$~\cite{lep-energy}.  
The complete error matrix from Reference~\citen{lep-energy} is used
to determine the uncertainties on $\MW$ and $\GW$ given in
Tables~\ref{tab:mw-syst} and~\ref{tab:gw-syst}, which are 
correlated between all final states.

As a cross check of the $\sqrt{s}$ calibration, events from the 
$\rm e^{+}e^{-} \rightarrow Z\gamma$ process with hard ISR were used 
to measure the mass of the Z boson~\cite{l3-277-z-return}. 
The Z-boson mass, $\MZ$, was determined to be 
\mbox{$91.272 \pm 0.032 \pm 0.033~\GeV$}, 
where the first uncertainty is statistical and the second systematic,
in agreement with the value measured at the 
Z resonance~\cite{l3-202-lineshape}, \mbox{$\MZ = 91.190 \pm 0.003~\GeV$}.
Assuming this value of $\MZ$, the method determines the 
average $\sqrt{s}$ to be \mbox{$175 \pm 68 \pm 68~\MeV$} lower than the 
value given by the LEP energy calibration, 
but consistent within the experimental uncertainty.

The intrinsic energy spread of the beams causes a $\sqrt{s}$ distribution
of the individual events with a Gaussian width of $240~\MeV$.
To assess this effect, the $\sqrt{s}$ constraint in the Monte Carlo events 
is varied by the same amount.
The changes of $\MW$ and $\GW$ are negligible.

\subsection{Lepton measurement}

The measurement of the lepton energy in $\QQEN$ and $\QQMN$ events
affects the mass reconstruction, while in the $\QQTN$ final state
it is solely based on the measurement of the jets.
Control samples of events from the $\EE\to\LL$ process are selected in 
calibration runs at the Z resonance and are used to cross check 
lepton reconstruction.  
The absolute energy scales for electrons and muons are known with a 
precision of $50~\MeV$.  
Varying the lepton energy scale by this amount and increasing the
lepton energy resolution in the simulation by 25\% of the value
measured with Z-resonance data, results
in the changes of $\MW$ and $\GW$ detailed in Table~\ref{tab:detector}.
Effects due to the determination of the lepton angles are negligible.  

\begin{figure}
  \includegraphics[width=0.49\textwidth]{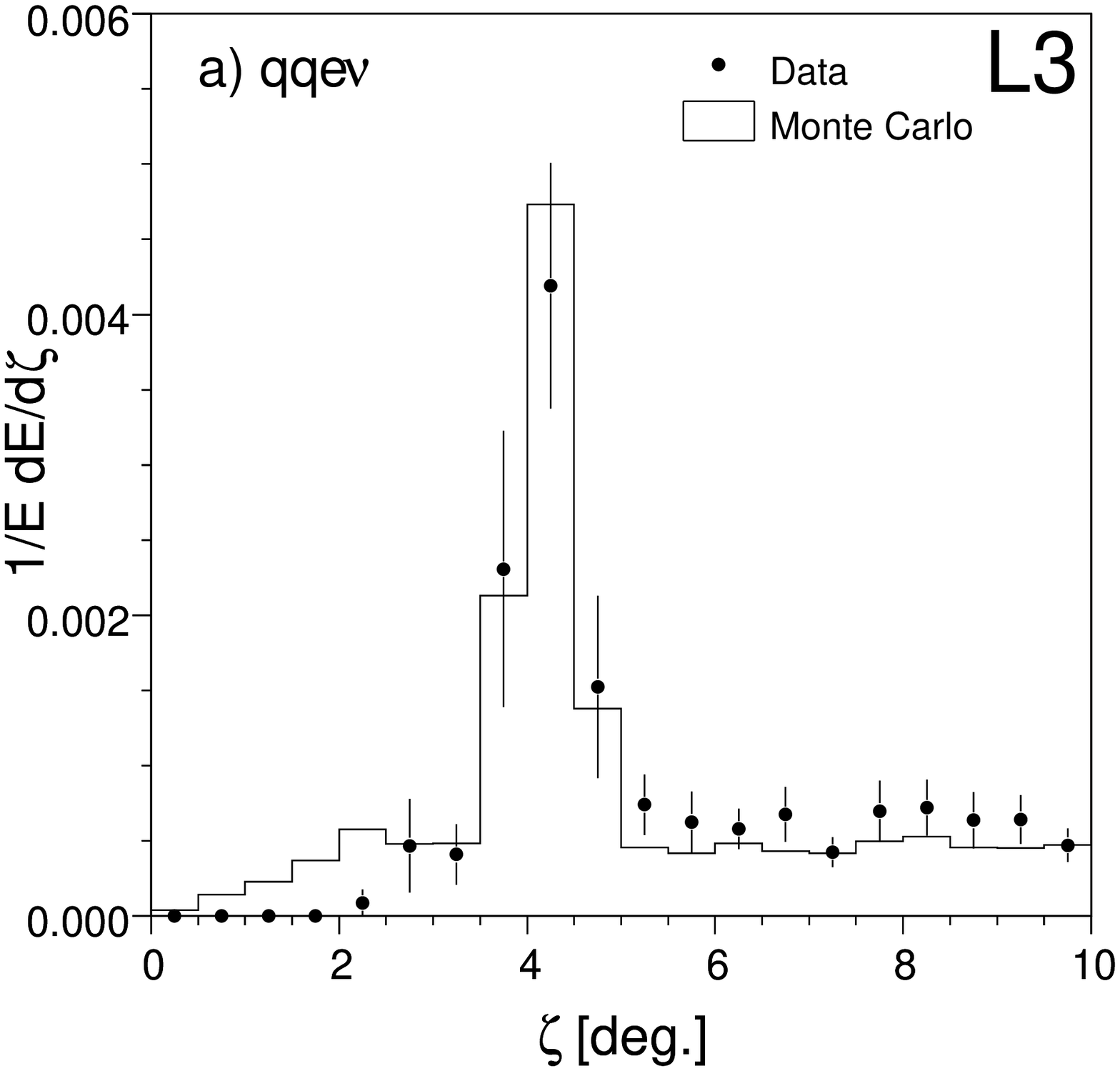}
  \hfill
  \includegraphics[width=0.49\textwidth]{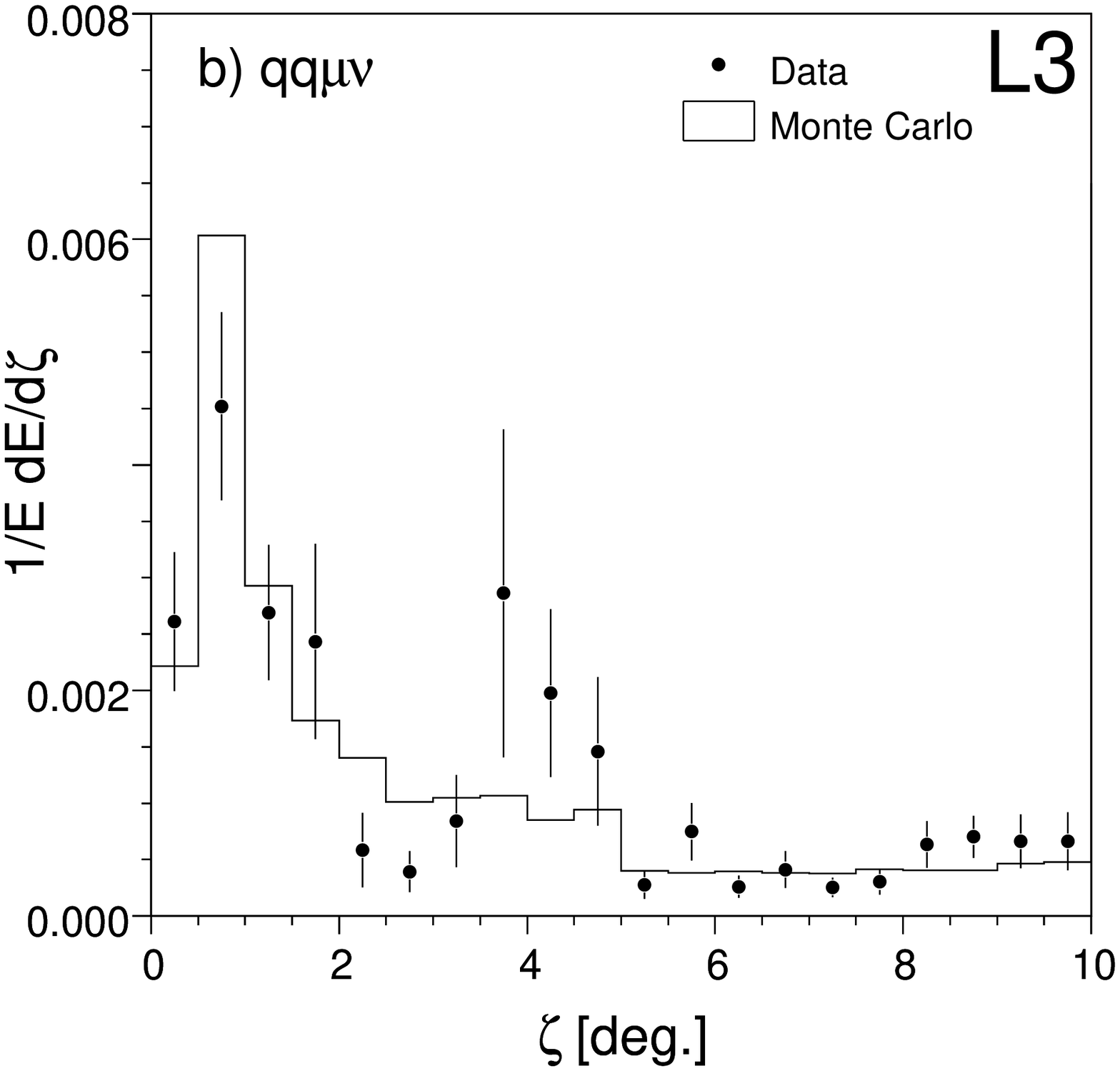}
  \caption{
    Calorimetric energy-flow versus the angle relative to the direction
    of the charged lepton, $\zeta$, 
    for a) the $\QQEN$ and b) the $\QQMN$ events.
    The error bars represent the standard deviation of the data
    distribution in each bin.
  }
\label{fig:lepcone}
\end{figure}

The distributions of the energy of calorimetric clusters around the charged 
lepton are shown in Figure~\ref{fig:lepcone}a and~\ref{fig:lepcone}b.
These clusters are normally joined to one of the jets.
If they are not correctly described by the Monte Carlo simulation,
this might result in a bias on the value of $\MW$. 
To assess this effect, all clusters within a cone of $5^\circ$ 
half-opening angle around the lepton are excluded from the jets.
No significant effect on $\MW$ is observed.

\subsection{Jet measurement}

The measurements of jet energies and directions affect the mass spectra
and are a potential source of systematic uncertainties on $\MW$ and $\GW$.
These uncertainties are assigned by varying the jet-energy scale by $50~\MeV$, 
smearing the jet energies by 1\% and smearing the jet directions by 
$0.5^\circ$.
The sizes of these variations correspond to the uncertainties estimated 
from $\EE\to\QQ$ events collected in calibration runs at the Z resonance.
These variations are applied to the Monte Carlo sample taken as reference to 
extract $\MW$ and $\GW$ from the data.
The effects on $\MW$ and $\GW$ are given in Table~\ref{tab:detector}.
As expected, the largest effect appears in the $\QQTN$ channel, 
where only the rescaled jets are used and no additional constraint is applied.

\begin{figure}
  \includegraphics[width=0.49\textwidth]{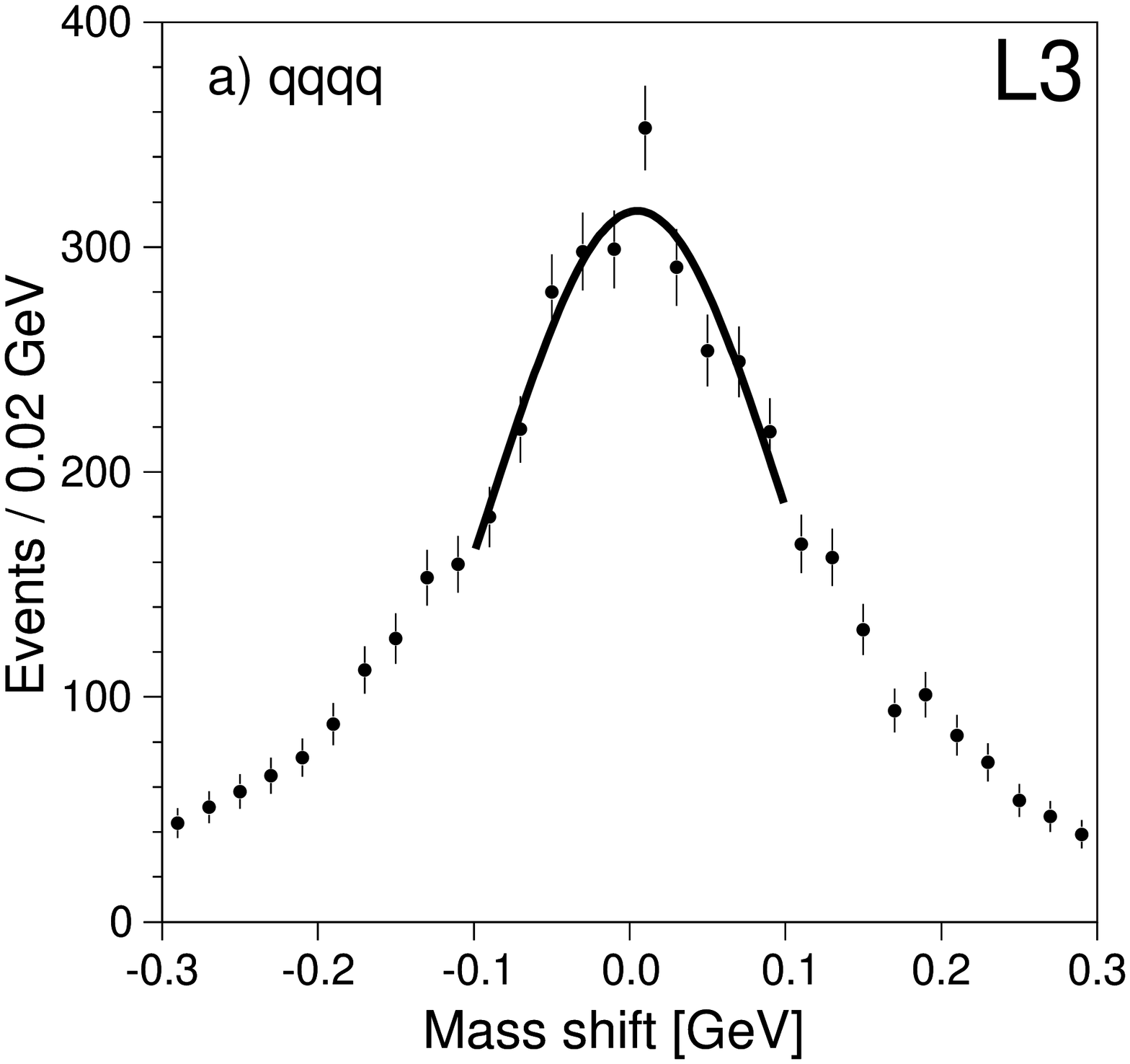}
  \hfill
  \includegraphics[width=0.49\textwidth]{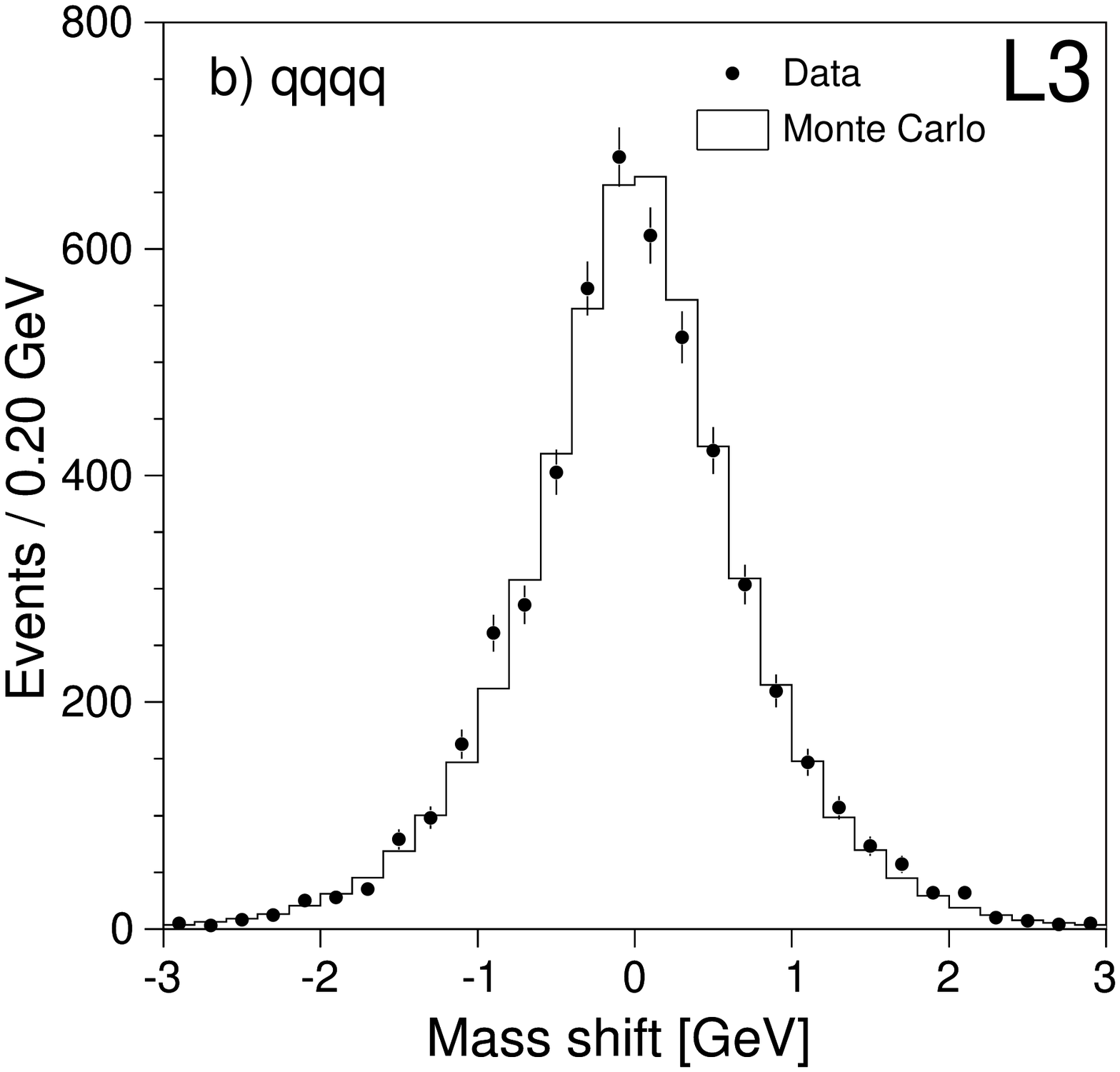}
  \caption{
    Distribution of the mass shifts between the standard analysis 
    of $\QQQQ$ events and analyses using 
    a) a displaced vertex and 
    b) jet reconstruction from tracking information only.
    A Gaussian fit is applied to the data distribution of a) and
    indicates an average mass shift consistent with zero, as shown by
    the curve.
    The data distribution of b) is in good agreement with the
    Monte Carlo prediction.
  }
\label{fig:angle-check}
\end{figure}

Generally, the event primary-vertex is shifted with respect to the
geometrical centre of the detector. If this shift was left uncorrected, it
would imply a systematic distortion of the jet angles. The actual position of the primary
vertex is measured using data and corrected for in the reconstruction
procedure. The shift is found to be less than 4~mm along the beam axis and
0.5~mm in the transverse plane, with an uncertainty of less than 5\%.
Figure~\ref{fig:angle-check}a shows for each data event the shift of the
reconstructed mass due to the vertex correction. Assuming that the vertex
is displaced within the uncertainty of its determination results in a
change of $\MW$ of less than $1~\MeV$.

Deviations of the calorimeter positions from their nominal locations 
would also cause angular distortions.
To check the angular measurement of the calorimetric clusters the measurement 
of $\MW$ is repeated using only clusters associated with tracks.
For each event, the mass is first reconstructed using the angular information
of the clusters and then from the angles of the associated tracks. 
These measurements are independent. 
The resulting mass-shift distribution is shown 
in Figure~\ref{fig:angle-check}b.
Combining all final states we obtain a change of $\MW$ of $-1\pm9~\MeV$ 
between both methods, consistent with zero.

Angle-dependent effects in the energy scale of the calorimeters could
lead to an additional bias in the measurement of the jet angles.
For instance, if forward clusters had a relative bias towards lower energy
than clusters in the central part of the detector, the direction of the
jet would be shifted towards the central detector region.
This effect is expected to be most evident in the $\QQQQ$ events,
which are strongly constrained by the kinematic fit.
To assess this effect the raw jet energies are compared to the jet energies 
after the kinematic fit for various polar-angle regions of the detector.
No significant change in $\MW$ is observed if the  cluster-energy scale in
the simulation is changed for each polar-angle region to agree exactly with
the data.

\begin{figure}
  \begin{center}
  \includegraphics[width=0.95\textwidth]{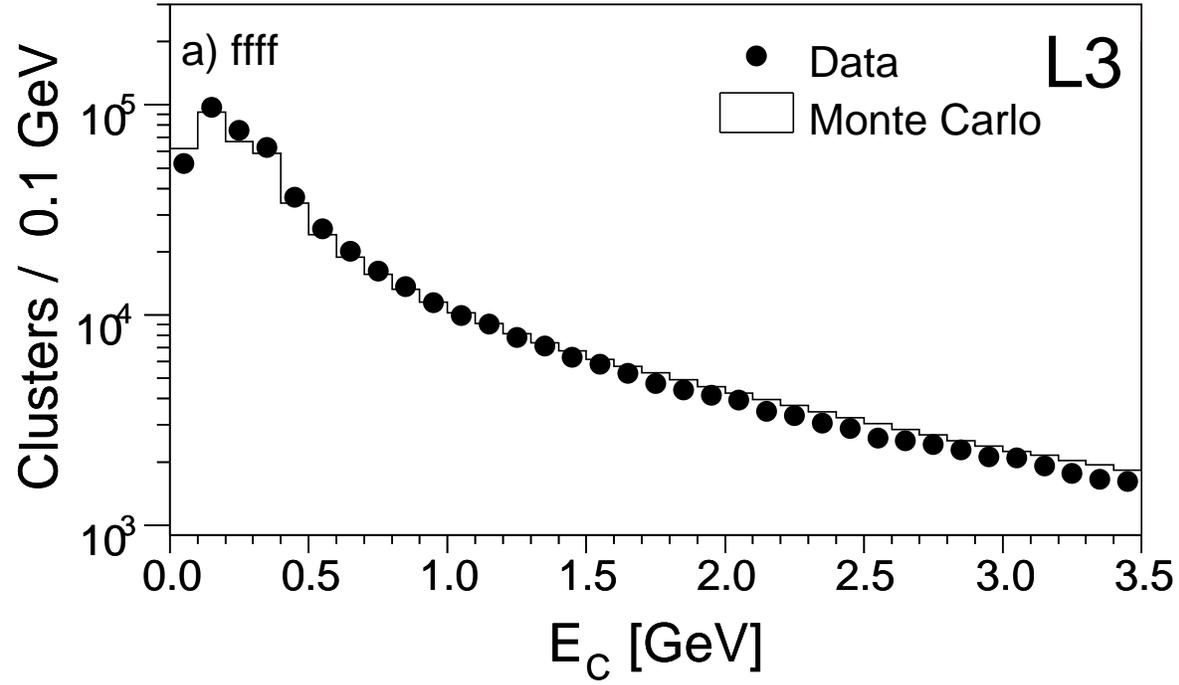}\vspace*{0.5cm}
  \includegraphics[width=0.95\textwidth]{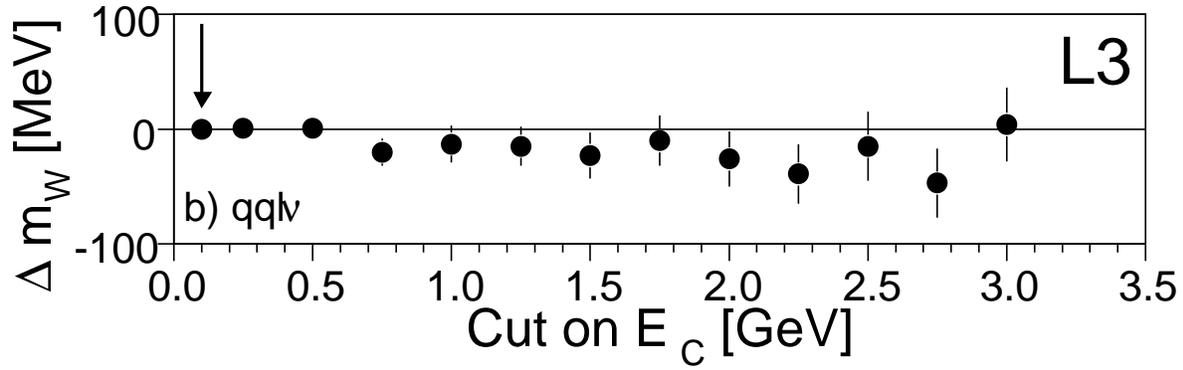}\vspace*{0.5cm}
  \includegraphics[width=0.95\textwidth]{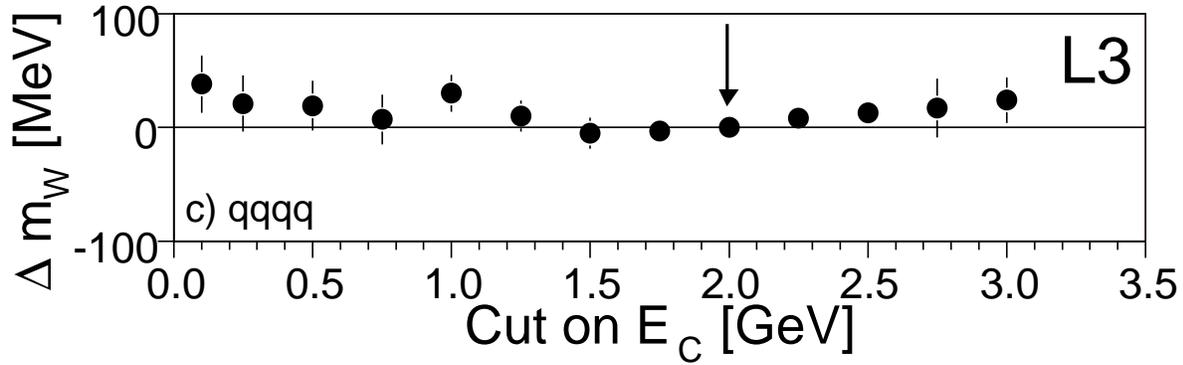}
  \end{center}
  \caption{
    a) Energy spectrum of the clusters used in the jet reconstruction
    and changes of $\MW$ 
    for b) the $\QQLN$ and c) the $\QQQQ$ final states
    caused by a variation of the cut on the minimum cluster energy, $E_C$.
    The arrows show the default values of the cut.
  }
\label{fig:ecut}
\end{figure}

\begin{figure}
  \begin{center}
  \includegraphics[width=0.95\textwidth]{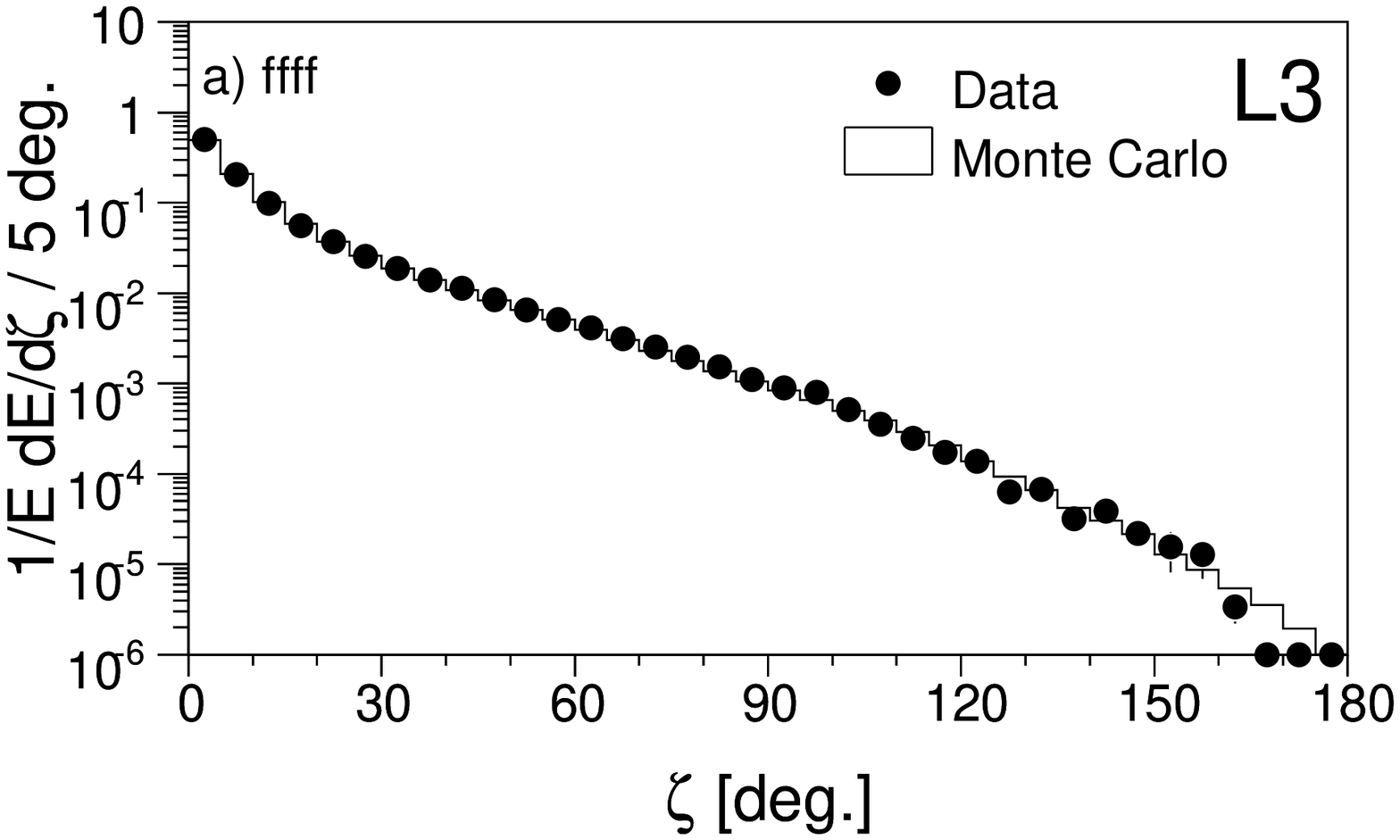}\vspace*{0.5cm}
  \includegraphics[width=0.95\textwidth]{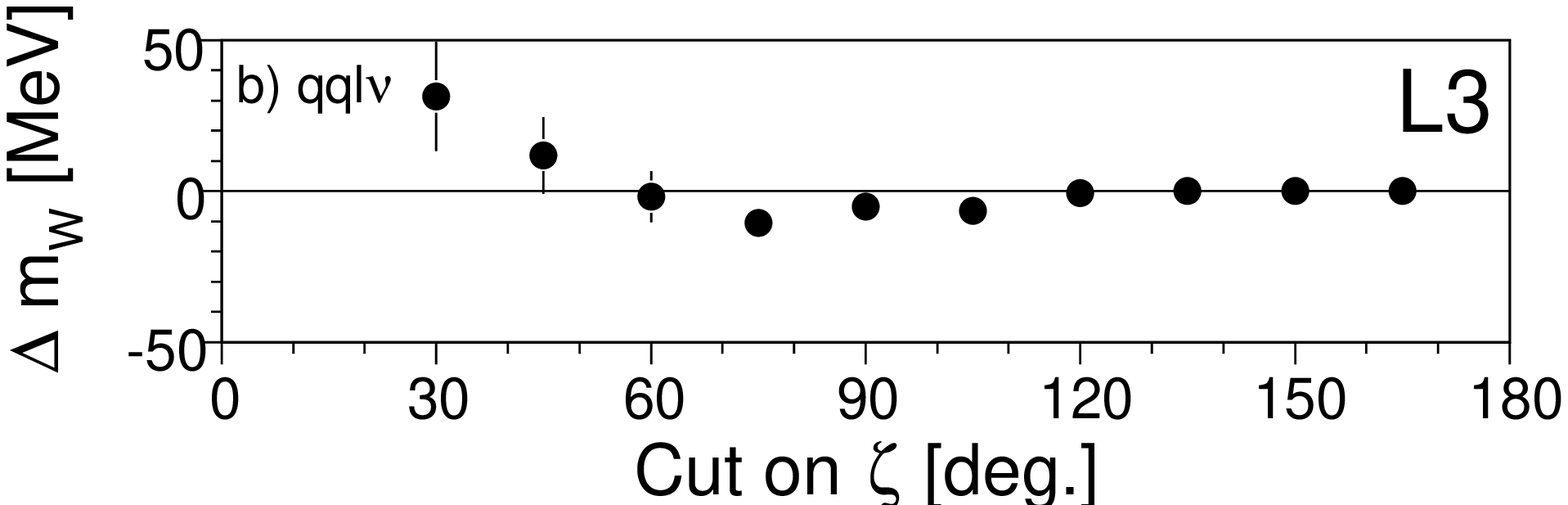}\vspace*{0.5cm}
  \includegraphics[width=0.95\textwidth]{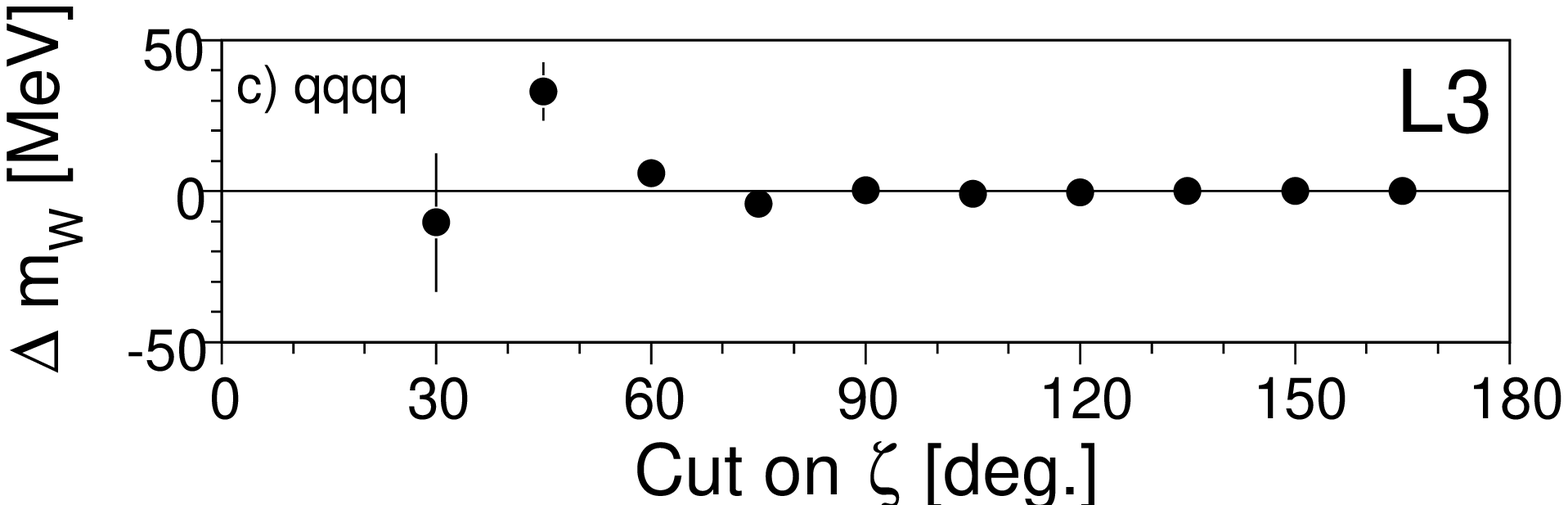}
  \end{center}
  \caption{
    a) Energy flow as a function of the angle relative to the jet direction, 
    $\zeta$,  and changes of $\MW$ 
    for b) the $\QQLN$ and c) the $\QQQQ$ final states
    after removing clusters outside a cone of 
    half-opening angle $\zeta$ around the jet direction.
  }
\label{fig:conecut}
\end{figure}

The energy spectrum of the clusters and the energy flow with respect to the 
jet axis are also investigated. 
They are shown in Figures~\ref{fig:ecut}a and \ref{fig:conecut}a, 
respectively.
Figures~\ref{fig:ecut}b and \ref{fig:ecut}c present the effect on $\MW$ 
when clusters below a given energy cut are removed. 
The changes of $\MW$ stay within the statistical uncertainty of the test
when varying the energy cut from the default values of $100~\MeV$ and $2~\GeV$
for $\QQLN$ and $\QQQQ$ events, respectively.
No significant change of $\MW$ is observed if clusters outside a cone around 
the jet axis are removed, as shown in Figures~\ref{fig:conecut}b
and \ref{fig:conecut}c for cones of half-opening angles from 
$30^\circ$ to $180^\circ$.

\subsection{Fit procedure}

The fit procedure determines $\MW$ and $\GW$ without
any bias as long as the Monte Carlo simulation  correctly describes effects such as
photon radiation and detector resolution.
This fit procedure is tested to high accuracy by fitting large Monte Carlo 
samples, typically a hundred times the size of the data sample.
The fits reproduce well the generated values $\MWgen$ and $\GWgen$
within the statistical accuracy of the test,
over a range of $\pm500~\MeV$ in $\MW$ and $\pm600~\MeV$ in $\GW$.

In the fit of $\MW$ the number of events per box is varied between 
350 and 450,
while in the fit of $\GW$ the minimal number of events is varied between 
150 and 250. 
In addition, the fit is restricted to masses in
the range between $70~\GeV$ and $90~\GeV$.
No statistically significant effect on $\MW$ or $\GW$ is observed
for any of these variations.

The reliability of the uncertainties given by the fit is tested by fitting 
for each final state several hundred small Monte Carlo samples, each the 
size of the data sample.  
The width of the distribution of the fitted central values agrees well with 
the mean of the distribution of the fit uncertainties. 

\subsection{Background}

Background which is not correctly described by the Monte Carlo simulation,
either in the total number of events or in their mass distribution,
could cause a shift of $\MW$ and $\GW$.
For both the semi-leptonic and the fully-hadronic selections, 
the four-fermion background is scaled by $\pm 5\%$. 
Additionally, background from the $\EE\to\QQ$ process
is scaled by $\pm 5\%$ and the slope of its mass spectrum is varied by 
$\pm10\%$ over the mass range between $65~\GeV$ and $95~\GeV$. 

The dominant background in the fully-hadronic selection is due to
$\EE\to\QQ$ events with multiple gluon radiation.
To better reproduce the four-jet rate observed in hadronic Z decays, 
a reweighting of the $\EE\to\QQ$ Monte Carlo events according to the 
value of $y_{34}$ is applied in our standard mass-extraction 
procedure~\cite{l3-291-ww-xsection}.
Removing this reweighting changes the total background contribution by 12\%
and shifts $\MW$ and $\GW$ by $10~\MeV$ and $80~\MeV$, respectively.
Half of the shift is taken as systematic uncertainty.
 
The effects on $\MW$ and $\GW$ due to the variation of height and
slope of the background mass spectrum and the uncertainty
due to the reweighting of the $y_{34}$ spectrum are
summarised in Table~\ref{tab:background}. 
The individual sources are added in quadrature to yield the systematic
uncertainty due to the background simulation.

\subsection{Monte Carlo statistics}

The systematic uncertainty due to the limited size of the signal Monte Carlo
sample used for the box fit is estimated by dividing it into several 
sub-samples of equal size and using each of them to fit the data.
The systematic uncertainty due to Monte Carlo statistics is then determined by
extrapolating the spread of the fit results to the full Monte Carlo sample.
The total systematic uncertainties on $\MW$ and $\GW$ due to limited 
Monte Carlo statistics are given for each final state in 
Tables~\ref{tab:mw-syst} and \ref{tab:gw-syst}, respectively.

\subsection{Photon radiation}

Four-fermion production, including its radiative corrections, is modelled
by the KANDY and RACOONWW Monte Carlo generators.
Both programs use pole expansions~\cite{pole-scheme} for the calculation
of ${\cal O}(\alpha)$ corrections.
KANDY models ISR using the Yennie-Frautschi-Suura (YFS) exponentiation 
scheme~\cite{YFS}, while FSR is simulated by the program PHOTOS~\cite{PHOTOS} 
in the case of charged leptons and by PYTHIA in the case of quarks.
Interference between ISR and FSR is neglected.
RACOONWW implements the full ${\cal O}(\alpha)$ matrix element for the 
radiative four-fermion production, $\EE \to \FFFF \gamma$.
Higher-order corrections coming from multiple ISR photons are implemented 
using a structure-function ansatz.
As the calculations implemented in RACOONWW are based on massless
fermions, the FSR simulation exhibits a minimum cut-off on the
photon-fermion angle.

The radiation of hard and isolated photons is better simulated by 
RACOONWW which implements the complete matrix element of the $\FFFF\gamma$ 
final state.
On the other hand, soft and collinear photons are not generated,
which makes the KANDY approach more appropriate for comparison with data.
KANDY uses a W propagator with a mass-dependent term containing the W width,
whereas RACOONWW uses a constant term.
Because the definitions differ by $27~\MeV$ in the position of the 
W peak~\cite{mw-def}, the $\MWgen$ input to RACOONWW is chosen $27~\MeV$ lower 
than for KANDY in order to give an identical W-boson lineshape.

A total of $300\, 000$ Monte Carlo events of the $\QQEN$, $\QQMN$ and
$\QQQQ$ final states are generated with the RACOONWW program 
at $\sqrt{s}=207~\GeV$, including full detector simulation.
Events with hard-photon radiation are selected at the generator level using 
the CALO5 algorithm~\cite{CALO5}, which recombines soft and collinear 
photons with the nearest fermion.
These events, after detector simulation, are used instead of data in
the mass fit which relies on KANDY 
as the reference Monte Carlo.
The change in $\MW$ from the comparison of the programs is derived and scaled
by the fraction of events with hard-photon radiation,
which is of the order of 10\%.
The same effect as observed in the $\QQMN$ channel is assumed for the
$\QQTN$ channel where no events were generated.
In an additional test, the KANDY events are reweighted such that
they represent the ${\cal O}(\alpha^2)$ ISR corrections instead of the 
${\cal O}(\alpha^3)$ calculation.
The changes of $\MW$ and $\GW$ resulting from these tests are detailed 
in Table~\ref{tab:photons}.
For each final state they are added in quadrature to estimate
the systematic uncertainties on $\MW$ and $\GW$ due to the modelling 
of photon radiation, given in Tables~\ref{tab:mw-syst} and~\ref{tab:gw-syst}.

\subsection{Hadronisation}

\begin{figure}
  \begin{center}
  \includegraphics[width=0.95\textwidth]{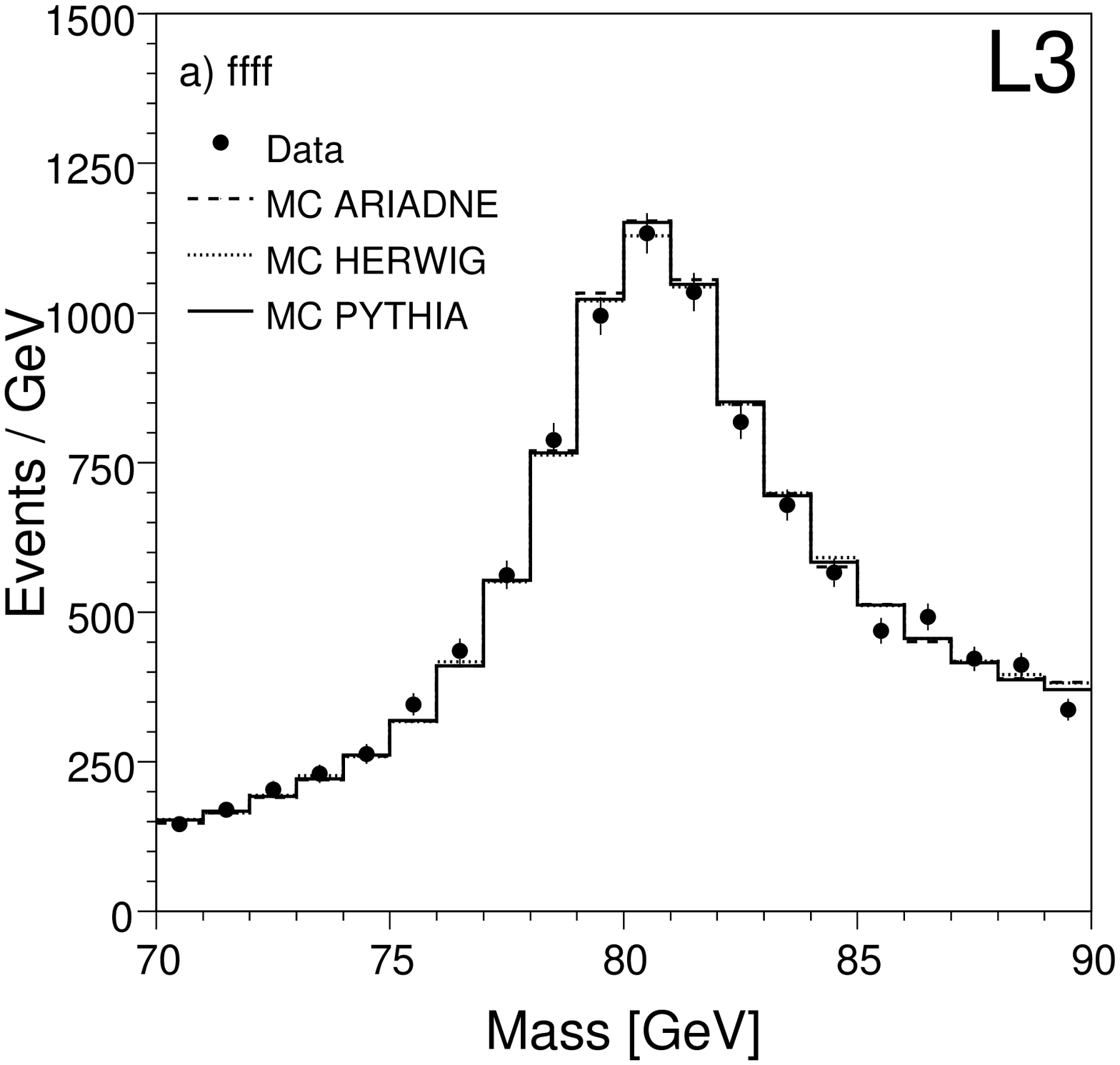}\vspace*{0.5cm}
  \includegraphics[width=0.95\textwidth]{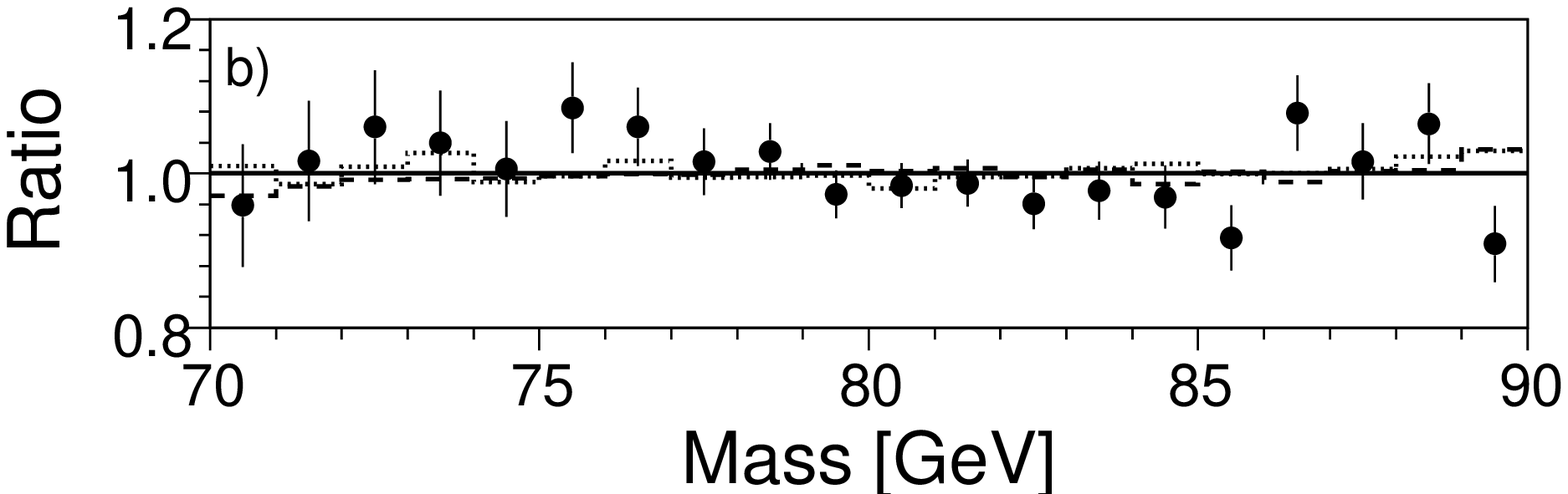}
  \end{center}
  \caption{
    a) Comparison of the reconstructed mass spectra,
    combined for all final states at $\sqrt{s} = 189~\GeV$,
    for data and for the three hadronisation models PYTHIA, ARIADNE and HERWIG 
    and b) the spectra normalised to the PYTHIA expectation.
  }
\label{fig:hadr-wmass}
\end{figure}

The hadronisation process is modelled by three different schemes
as implemented in the Monte Carlo programs PYTHIA, HERWIG and ARIADNE.
For the perturbative phase PYTHIA and HERWIG simulate a parton shower, 
while a dipole cascade is produced in ARIADNE.
The Lund string-hadronisation model is used by PYTHIA and ARIADNE,
while HERWIG employs a cluster model.
A comparison of the mass distributions of the three different models 
with data is shown in Figure~\ref{fig:hadr-wmass}.
Within the statistical accuracy, all three Monte Carlo distributions are 
compatible with the data.

The results for $\MW$ and $\GW$ presented in this paper are based on the 
PYTHIA model.
Systematic effects due to modelling of the hadronisation process are 
determined by comparison with the other two programs.
In the mass-extraction fit the data events are replaced by high-statistics 
samples of Monte Carlo events generated with PYTHIA, HERWIG and ARIADNE.
These Monte Carlo samples consist of events which are identical at the
four-fermion level and thus differ only in their hadronisation.
The changes of $\MW$ and $\GW$ due to the use of HERWIG or ARIADNE
are listed in Table~\ref{tab:hadronisation}.
For $\MW$, HERWIG and ARIADNE reproduce the PYTHIA results within 
the statistical uncertainty, except in the $\QQTN$ channel. 
This is mainly caused by the misassignment of energy deposits from 
the remainder of the event to the tau-lepton jet which is based on
a cone definition.
This effect, altering the reconstruction of the jets and therefore $\MW$, 
strongly depends on the choice of the hadronisation model.
For $\GW$, ARIADNE is in good agreement with PYTHIA, while HERWIG shows 
significant differences, especially for semi-leptonic final states.  

\begin{figure}
  \begin{center}
  \includegraphics[width=0.95\textwidth]{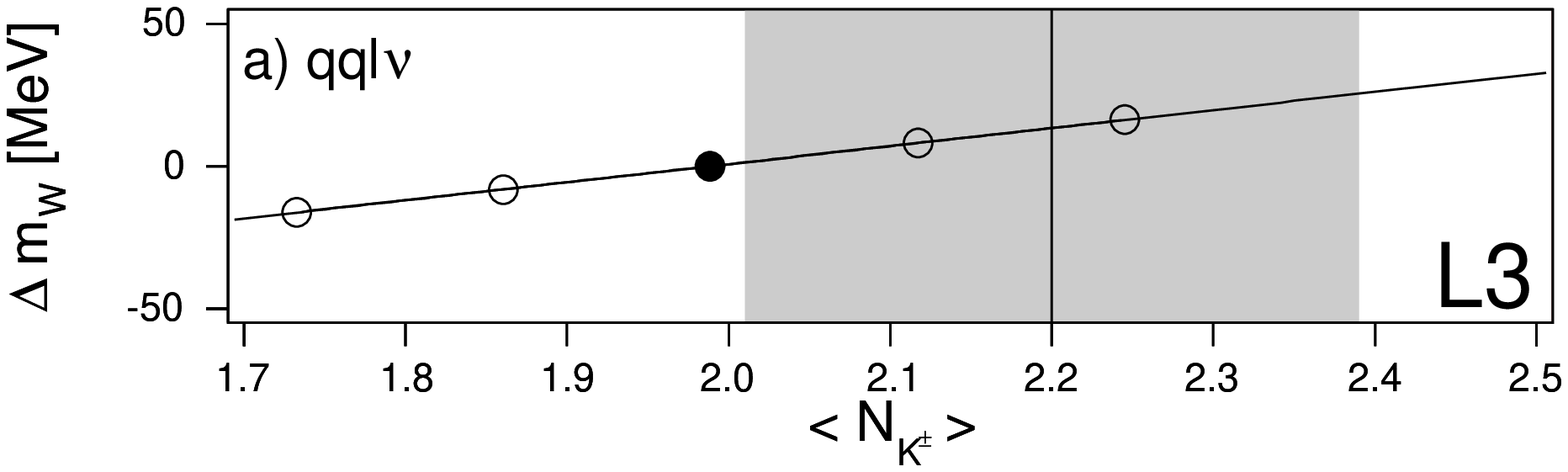}
  \includegraphics[width=0.95\textwidth]{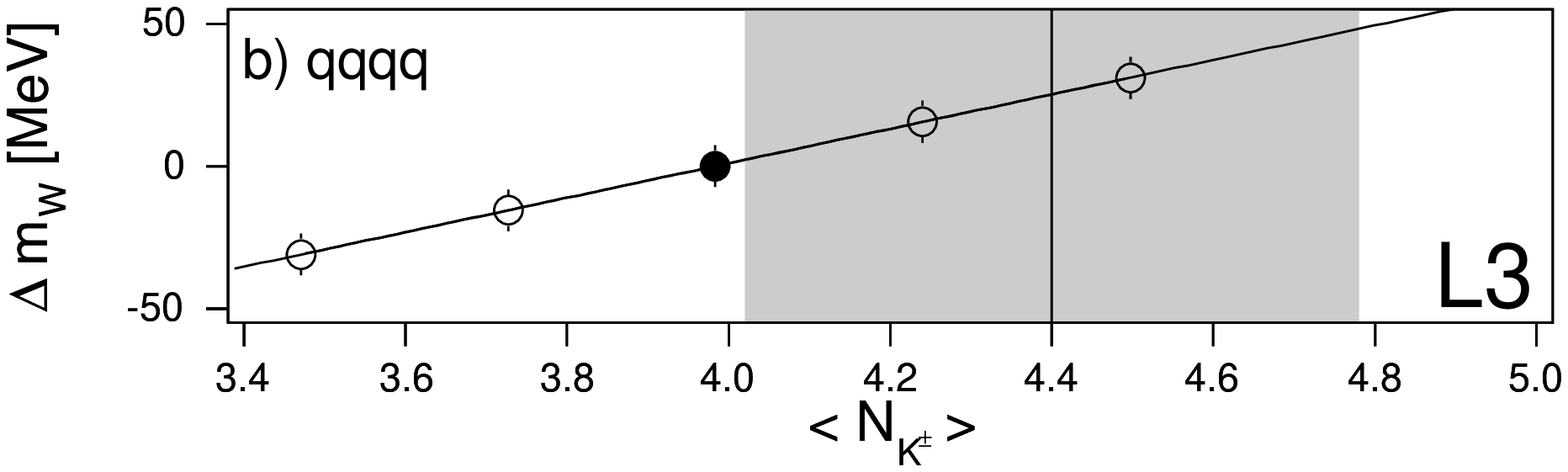}
  \includegraphics[width=0.95\textwidth]{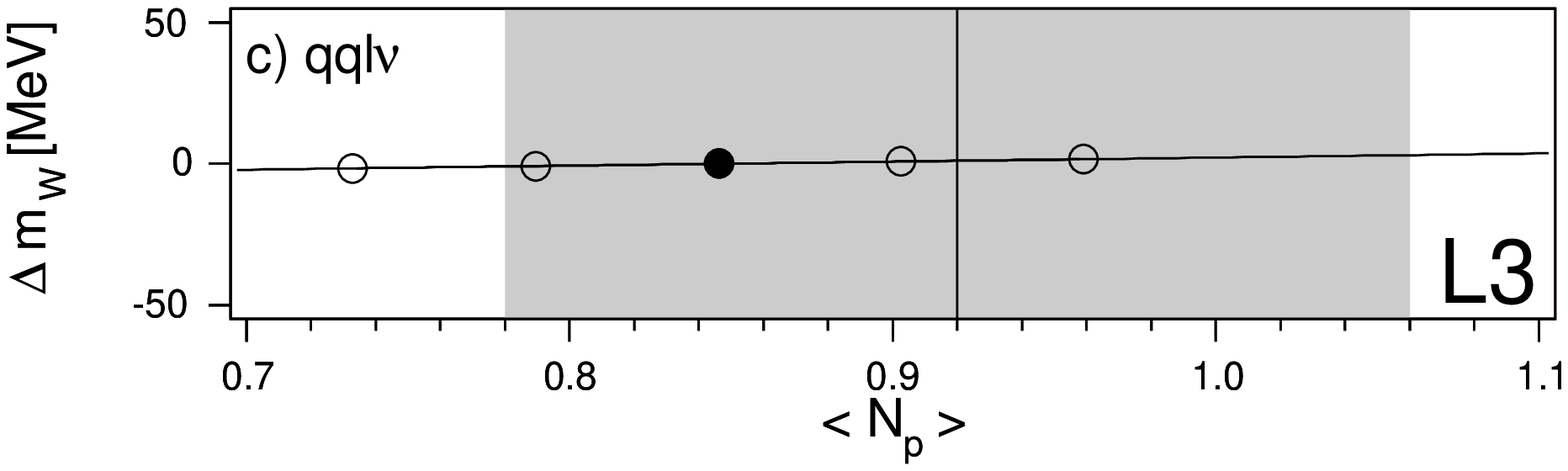}
  \includegraphics[width=0.95\textwidth]{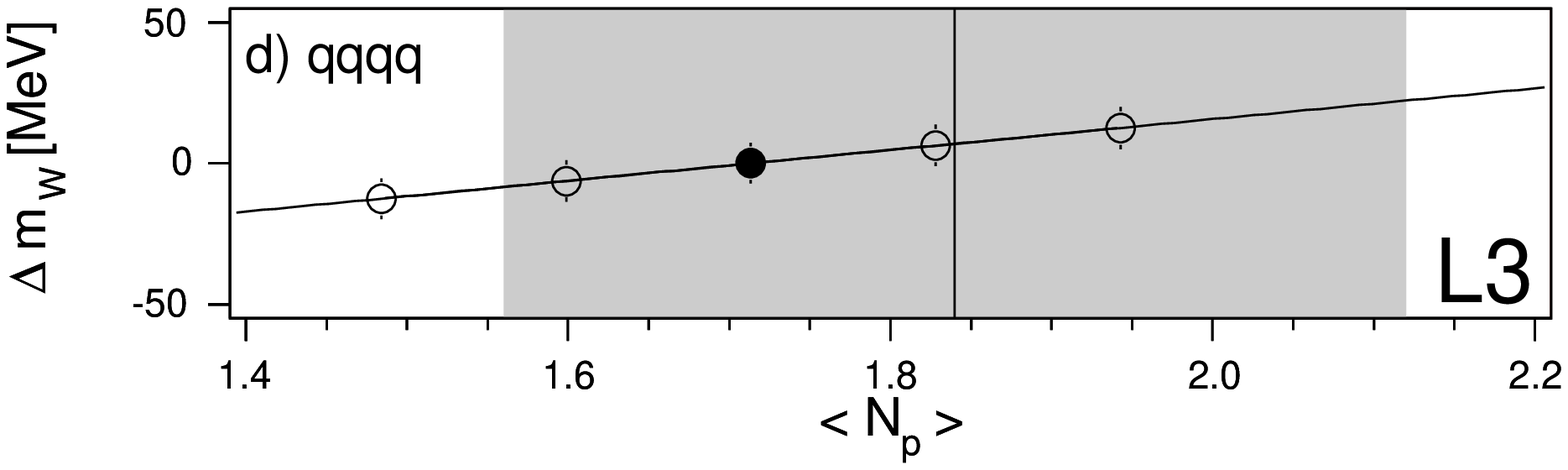}
  \end{center}
  \caption{
    Changes of $\MW$ due to reweighting Monte Carlo events according
    to  the mean charged-kaon multiplicity for 
    a) the $\QQLN$ and b) the $\QQQQ$ events and 
    of the mean proton multiplicity for 
    c) the $\QQLN$ and d) the $\QQQQ$ events.
    The full circles show the default values of our simulation whereas 
    the vertical lines show the measured multiplicities and the grey bands 
    their uncertainties~\cite{delphi-kaon-proton}.
  }
\label{fig:reweighting}
\end{figure}

The four-momenta of calorimetric clusters which are used to form hadronic jets
are calculated using the energy and angle measurements and assuming their 
masses to be either zero or the pion mass.
However, kaons and protons are frequently produced resulting in a shift
of the jet masses.
This shift is automatically corrected in the mass-extraction fit 
which uses the Monte Carlo simulation containing these hadrons.
If the simulation predicts different multiplicities for these heavier 
hadrons than is present in data, systematic effects on the measurement of
$\MW$ and $\GW$ are expected. 
In order to assess this systematic effect,
the mean number of charged kaons and protons produced in the W-boson decays
of our simulation is compared to
measurements~\cite{delphi-kaon-proton} and found to be in agreement.
The shifts $\Delta\MW$ and $\Delta\GW$ are calculated with
Monte Carlo events reweighted such that the mean kaon and proton 
multiplicities agree exactly with the measured values.
It is checked that the mass spectrum at generator level is not 
distorted by this reweighting procedure.
Figure~\ref{fig:reweighting} shows the linear dependence of $\Delta\MW$
on the average kaon and proton multiplicity.
This linear dependence is used to translate the uncertainty of the measured 
kaon and proton multiplicities into uncertainties on $\MW$ and $\GW$.
Table~\ref{tab:reweighting} presents the shifts and uncertainties of 
$\MW$ and $\GW$ due to the correction of the Monte Carlo simulation.

In the $\QQQQ$ final state, the DURHAM parameter $y_{45}$ is used to 
discriminate events with hard-gluon radiation.
This variable might be affected by hadronisation uncertainties.
A change of the selection criterion $\log y_{45}<-6.2$ between 
$-5.8$ and $-6.6$ has no significant influence on $\MW$.

The average absolute shift of $\MW$ and $\GW$ due to the alternative 
hadronisation models ARIADNE and HERWIG and the uncertainty deduced from 
the variation of the kaon and proton multiplicities in the Monte Carlo 
simulation are added in quadrature
to yield the total uncertainties due to the hadronisation modelling,
given in Tables~\ref{tab:mw-syst} and \ref{tab:gw-syst}.

\subsection{Final state interactions in fully-hadronic events}
\label{sec:syst-fsi}

The Monte Carlo programs hadronise the quarks from the two
W bosons independently.
However, CR effects would invalidate this assumption and thus 
affect the mass reconstruction.
Similarly, BEC between bosons arising from different W bosons,
if incorrectly modelled, could have the same effect.

Our measurements~\cite{l3-174-ZBE,l3-242-ZBE,l3-257-BE} of BEC
indicate that correlations in hadronic W-boson decays are very similar 
to those in Z-boson decays into light quarks.  
Furthermore, BEC between hadrons from different W bosons are disfavoured.
They are limited to at most 30\% of the strength simulated in the BE32 
model~\cite{MC-BE} implemented in PYTHIA~5.7.  
Since all our previous mass measurements at $\sqrt{s}=172-183~\GeV$ were
performed under the assumption of full inter-W BEC, the results 
obtained in the $\QQQQ$ channel are re-evaluated in light of our measurement
of vanishing inter-W BEC.  

\begin{figure}
  \centerline{\includegraphics[width=0.95\textwidth]{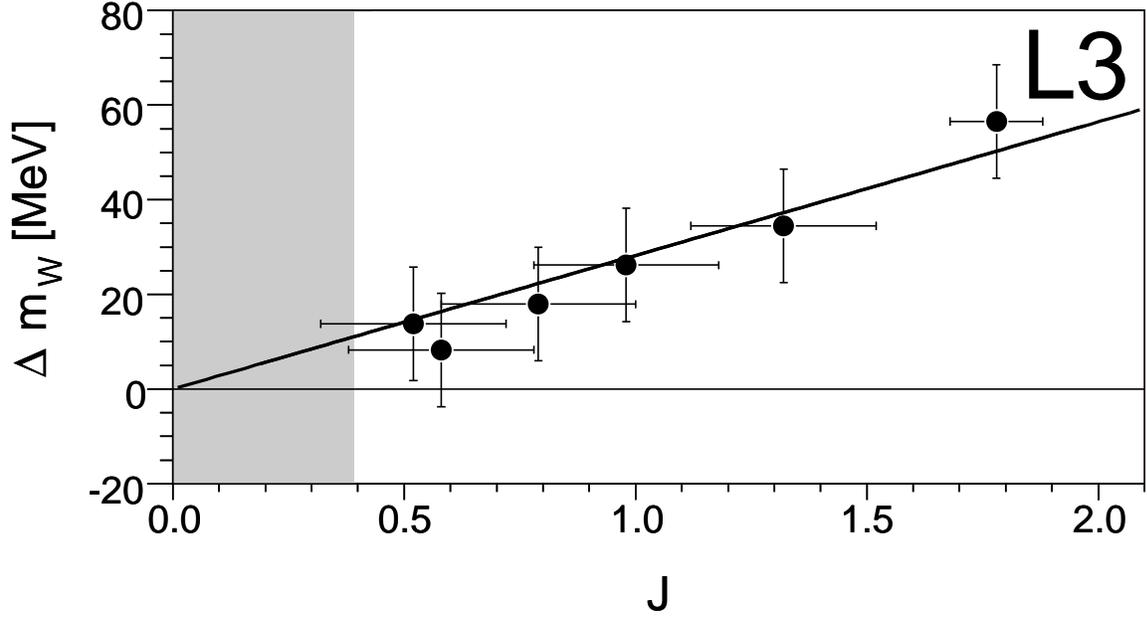}}
  \caption{
    Changes of $\MW$ with respect to the observable $J$ for
    Monte Carlo samples of the BE32 model with different BEC parameters
    at $\sqrt{s} = 189~\GeV$.
    The grey band shows the range of $J$ which is compatible with our
    BEC measurement~\cite{l3-257-BE} at the 68\% confidence level.
  }
\label{fig:BE-linearity}
\end{figure}

Reference~\citen{l3-257-BE} presents the L3 measurement of the difference
between the two-particle densities of the data and the 
simulation without inter-W BEC, $\Delta\rho_2(Q)$.
The integral, $J$, of this difference is measured to be below $0.39$
at 68\% confidence level.
For different Monte Carlo samples, generated with various strengths of inter-W
BEC, but fixed strength of the intra-W BEC, the integral $J$ is determined.
The shift $\Delta\MW$ exhibits a linear dependence with respect to $J$,
as shown in Figure~\ref{fig:BE-linearity}.
The effects on $\MW$ and $\GW$ for a maximum inter-W BEC, as allowed by our
direct BEC measurement, are detailed in Table~\ref{tab:BE}. A linear
dependence of BEC effects on $\sqrt{s}$ is assumed.

\begin{figure}
  \centerline{\includegraphics[width=0.95\textwidth]{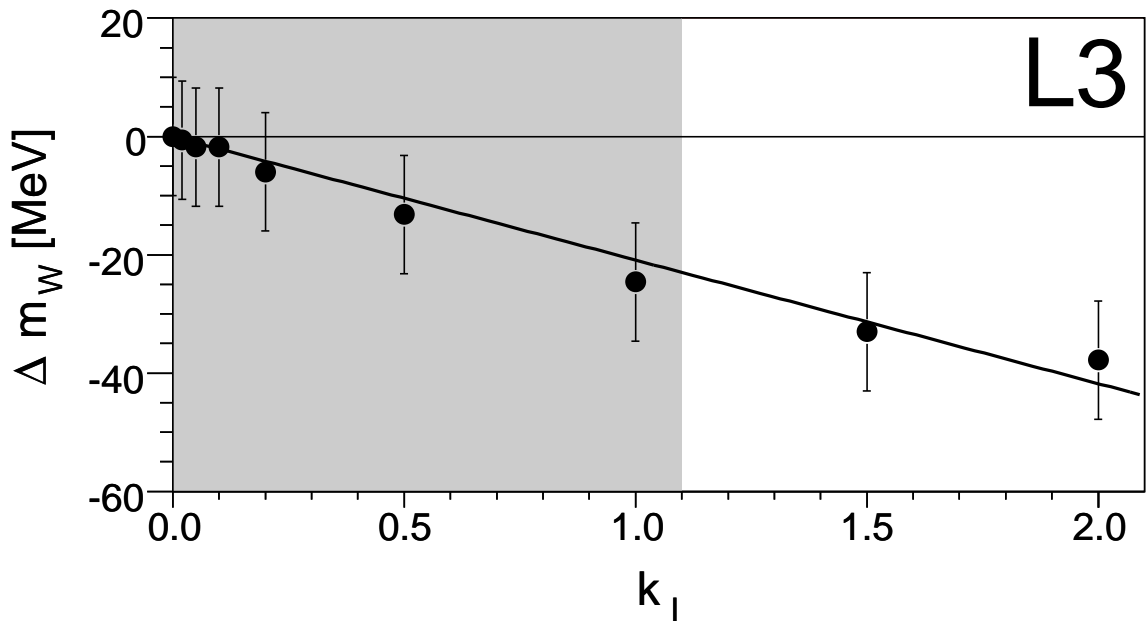}}
  \caption{
    Changes of $\MW$ with respect to the parameter $k_\mathrm{I}$ 
    of the SK-I model at $\sqrt{s}=189~\GeV$.
    The cut on the minimum cluster energy of $2~\GeV$ is applied.
    The grey band shows the range of $k_\mathrm{I}$ which is compatible 
    with our CR measurement~\cite{l3-266-CR} at the 68\% confidence level.
  }
\label{fig:CR-kappa}
\end{figure}

A dedicated study of reconnection effects in the particle flow between jets 
in $\QQQQ$ events shows that the data are consistent with no or only a small 
CR effect~\cite{l3-266-CR}. 
A 68\% upper limit on the CR parameter $k_I$ is set at $1.1$ in the framework 
of the SK-I model~\cite{MC-CR-SK} as implemented in PYTHIA~5.7. 
The influence of the CR parameter $k_\mathrm{I}$ on $\MW$ is studied
in the SK-I framework by mixing event samples simulated at $\sqrt{s}=189~\GeV$ 
with full and without inter-W CR.
The result is shown in Figure~\ref{fig:CR-kappa} for moderate values of 
$k_\mathrm{I}$ where a linear dependence can be assumed. 
The particle flow analysis is found to be insensitive to CR effects 
implemented in other models such as ARIADNE type~II~\cite{MC-CR-ARIADNE} 
and HERWIG~\cite{MC-CR-HERWIG}.
The ARIADNE-II model is compared to the ARIADNE-I model, the latter having
been modified such that in both models the shower cascade is performed in two 
phases with an identical cut-off parameter.

These Monte Carlo studies show that the effect of CR on $\MW$ grows with
increasing $\sqrt{s}$ in the case of the SK-I model, while only little 
dependence on $\sqrt{s}$ is seen for ARIADNE and HERWIG.
For all energies and all models the shift of $\MW$ is comparable or
smaller than the shift predicted by the SK-I model at $k_\mathrm{I} = 1.1$, 
which is used to estimate the systematic uncertainty due to CR effects.
It is interesting to note that studies of the distribution of particles
in the inter-jet region of three-jet hadronic Z decays exclude the
predictions of the CR models of ARIADNE and HERWIG for this 
case~\cite{l3-272-3jetBE}.
No version of the SK-I model applicable to Z decays exists.

\begin{figure}
  \centerline{\includegraphics[width=0.95\textwidth]{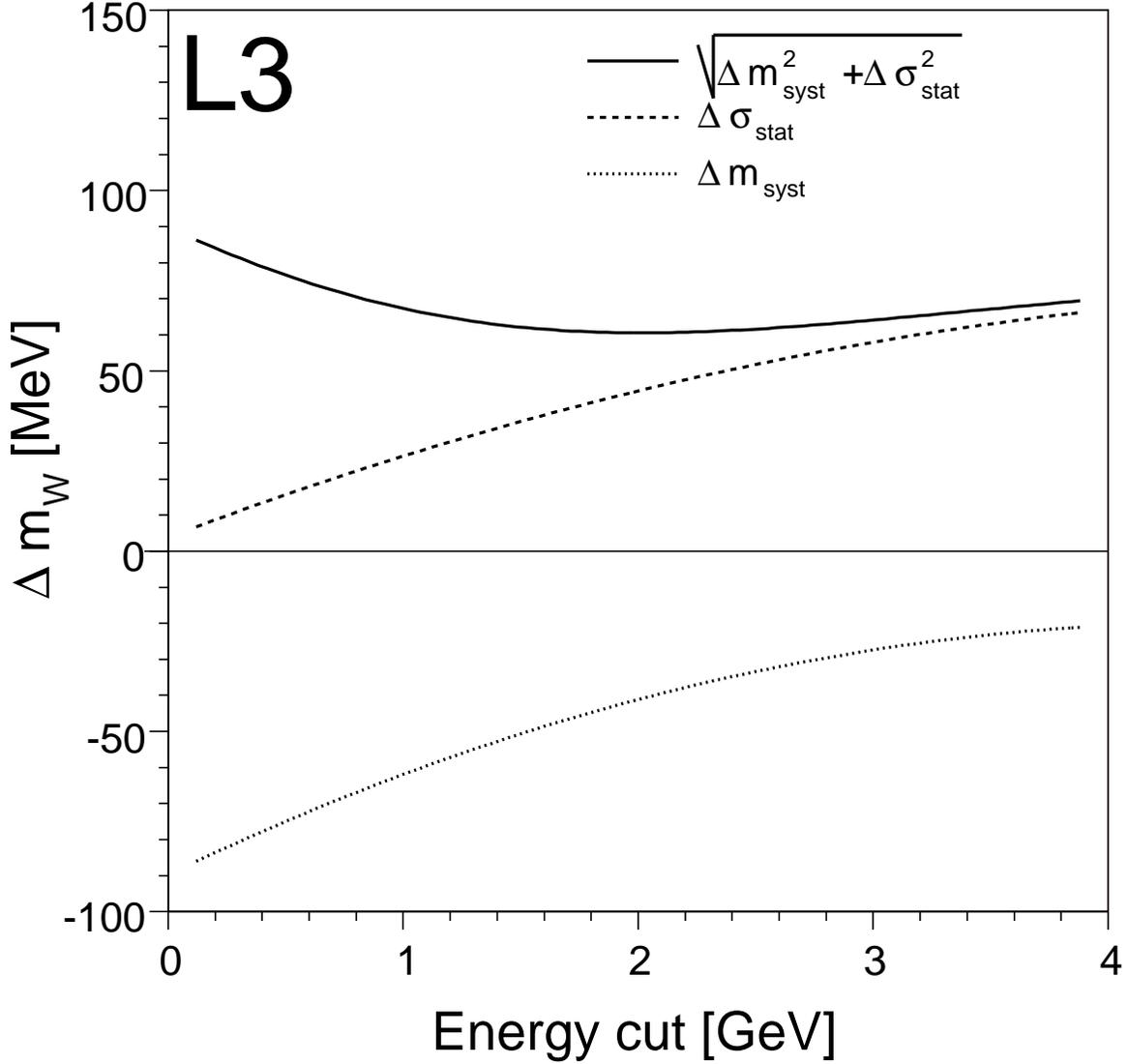}}
  \caption{
    CR effects simulated with the Monte Carlo model SK-I
    calculated after removing clusters with an energy below a given 
    threshold energy.
    The change of the final $\MW$ measurement in the $\QQQQ$
    channel, $\Delta m_\mathrm{syst}$, when the default simulation 
    without CR is replaced by the SK-I model using $k_\mathrm{I} =
    1.1$ is shown.
    The additional component of the statistical uncertainty on the
    final $\MW$ result, $\Delta \sigma_\mathrm{stat}$,  after applying 
    the given energy cut and the quadratic sum of both effects is also shown.
  }
\label{fig:CR-cut-cone}
\end{figure}

The use of a cone algorithm for jet clustering lowers the sensitivity
to CR effects, as the analysis is less affected by the inter-jet regions 
where the influence of CR is largest.
More effectively, removing clusters below a certain energy cut rejects
particles predominantly produced during the non-perturbative phase of the
hadronisation process where CR effects take place.
Monte Carlo studies are performed at $\sqrt{s}=189~\GeV$ applying various 
cuts on the minimum cluster energy.
The dependence of the $\MW$ shift on the energy cut is extrapolated 
to the full data sample and shown in Figure~\ref{fig:CR-cut-cone}.
The additional component to the statistical uncertainty due
to the slight degradation of the mass resolution caused by the cut
is calculated and added in quadrature to the shift of $\MW$.
A cut at minimum cluster energies of $2~\GeV$ is found to be the 
optimal choice and is therefore used in the extraction of $\MW$ and $\GW$
from the data of the $\QQQQ$ final state.
Table~\ref{tab:CR} presents the effect of CR on $\MW$ and $\GW$.

Monte Carlo studies show that the relative reduction of the $\MW$ shift
due to the energy cut is independent of $k_\mathrm{I}$ and $\sqrt{s}$.
The mass shifts observed for the SK-I Monte Carlo simulation with full CR at various 
$\sqrt{s}$ values are obtained using the dependence on $k_\mathrm{I}$ 
and on the energy cut extracted at $\sqrt{s}=189~\GeV$. 
The systematic uncertainties on $\MW$ are calculated 
using a linear dependence on $\sqrt{s}$ and assumed to be fully correlated.
For $\GW$ no $\sqrt{s}$ dependence is seen.

\section{Results}
\label{sec:results}

\begin{figure}
  \centerline{\includegraphics[width=0.95\textwidth]{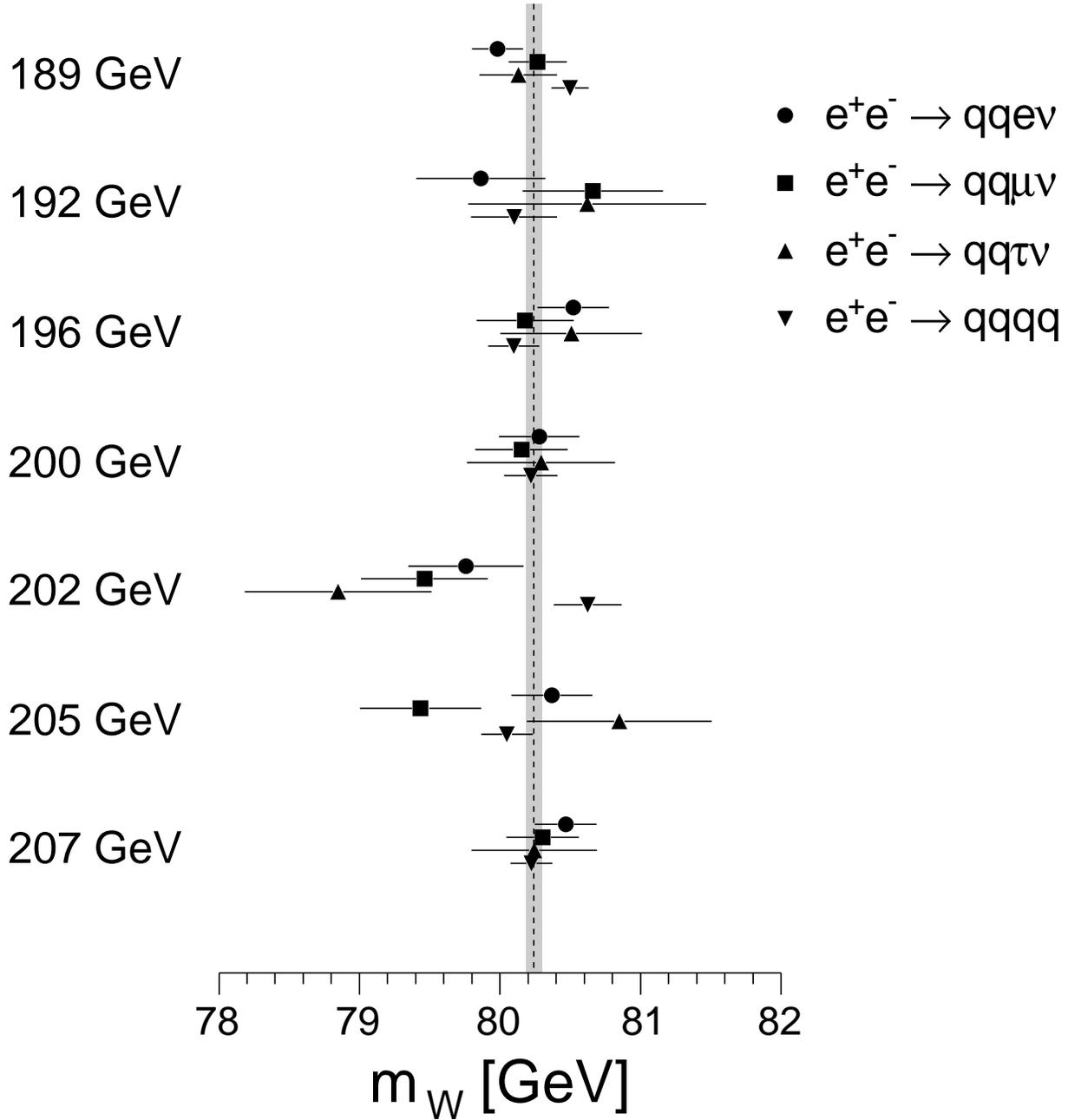}}
  \caption{
    The results for $\MW$ for the four final states
    and the seven average $\sqrt{s}$ values. Statistical and
    systematic uncertainties are added in quadrature.
    The combined $\MW$ result and its uncertainty are
    indicated as the dashed line and the grey band, respectively.
  }
\label{fig:mw-all}
\end{figure}

Figure~\ref{fig:mw-all} compares the $\MW$ measurements in the four
different final states at the seven average $\sqrt{s}$ values.
The measurements of $\MW$ and $\GW$ from the individual final states
are combined using the ``best linear unbiased estimate'' technique~\cite{lyons}.
This combination method takes into account all systematic
uncertainties as well as their correlations. 
When combining measurements taken at different $\sqrt{s}$ values, the
correlations of the LEP energy determination~\cite{lep-energy} are used. 
Within each final state, the uncertainties due to lepton measurement,
background determination, BEC and CR are taken as fully correlated
between the measurements at different $\sqrt{s}$. 
The uncertainties due to jet measurement, photon radiation
and hadronisation are fully correlated between all final states and between 
all $\sqrt{s}$ values. 
The systematic uncertainty due to limited Monte Carlo statistics remains 
uncorrelated for all measurements.
In the case of the simultaneous estimate of $\MW$ and $\GW$, the correlations 
between both parameters as determined in the individual box fits
are included in the combination procedure. 

\begin{figure}
  \centerline{\includegraphics[width=0.95\textwidth]{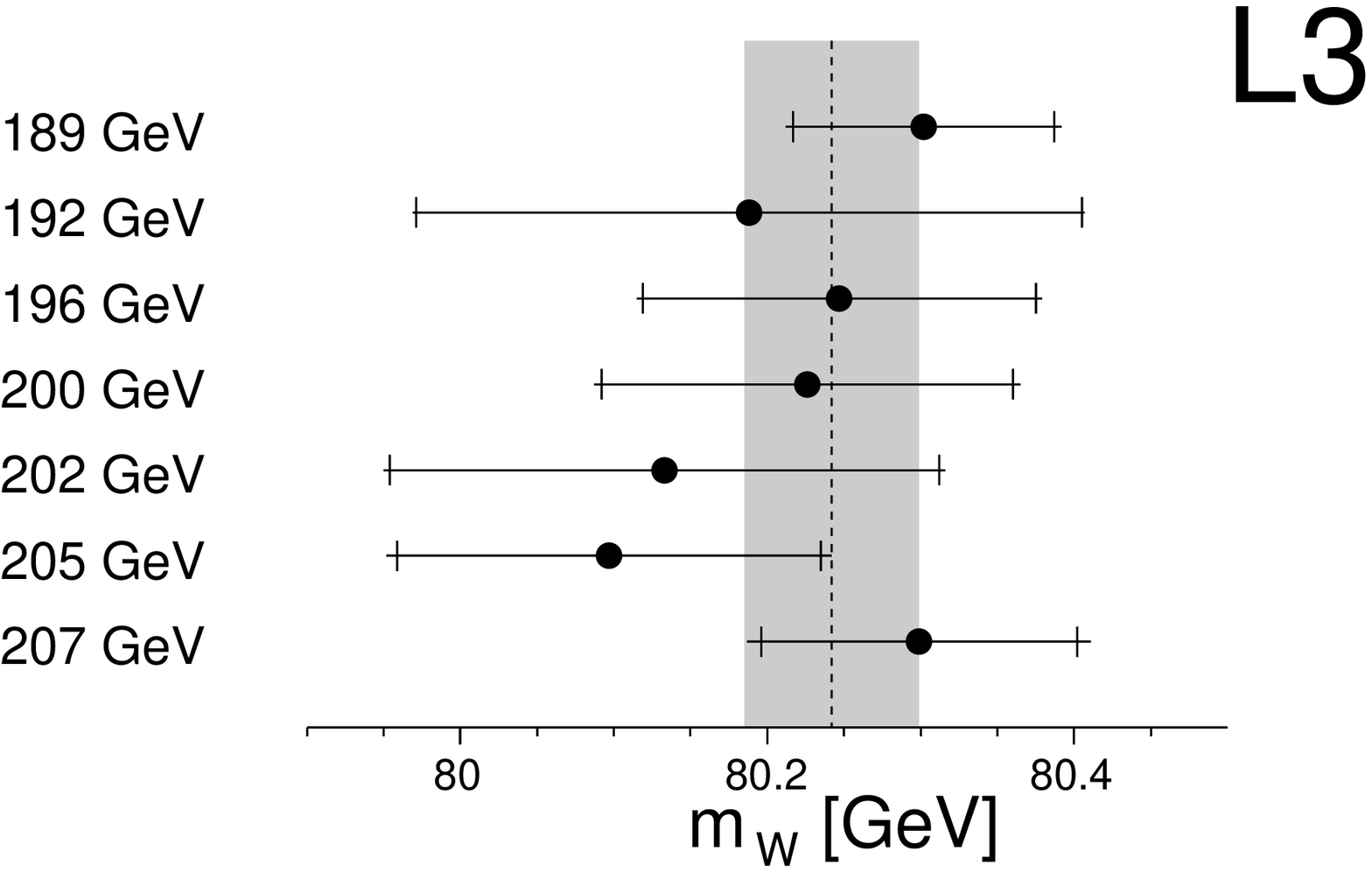}}
  \caption{
    Comparison of the results for $\MW$ for the seven average $\sqrt{s}$ values.
    The inner error bar represents the statistical uncertainty.
    The combined $\MW$ result and its uncertainty are
    indicated as the dashed line and the grey band, respectively.
  }
\label{fig:mw-sqrts}
\end{figure}

\begin{figure}
  \centerline{\includegraphics[width=0.95\textwidth]{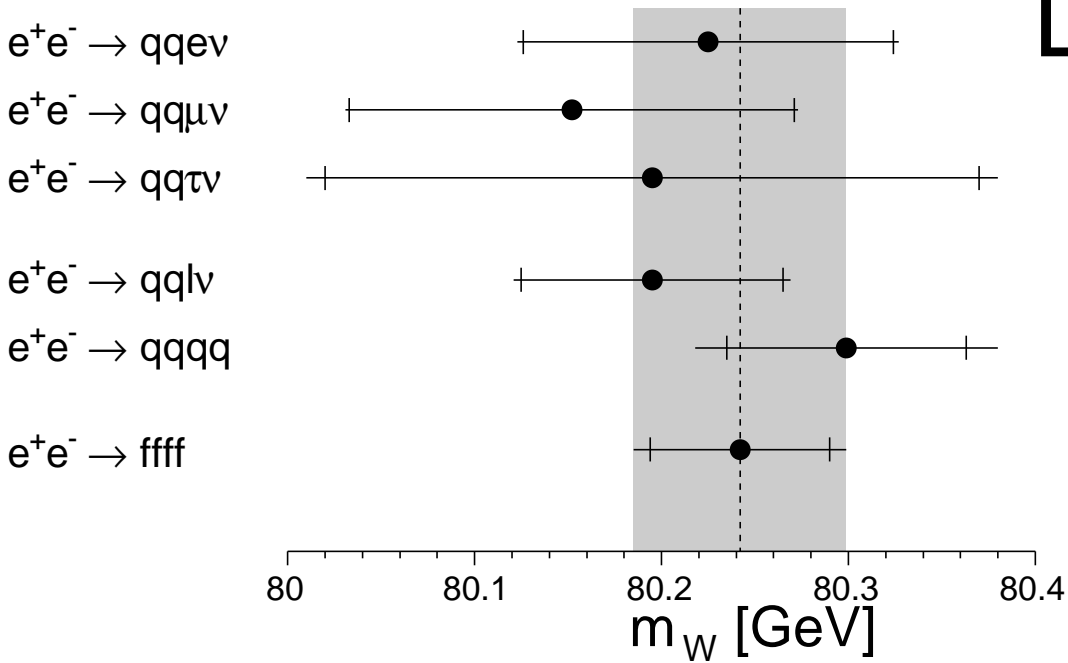}}
  \caption{
    Comparison of the results for $\MW$ for the $\sqrt{s} = 189 - 209~\GeV$
    in each of the different final states.
    The inner error bar represents the statistical uncertainty.
  }
\label{fig:mw-channel}
\end{figure}

Combined results of $\MW$ are shown in Figure~\ref{fig:mw-sqrts}
for each $\sqrt{s}$ value averaged over the final states. 
Figure~\ref{fig:mw-channel} shows the results for each final state and
their combination.
Table~\ref{tab:mw-results} gives the results on $\MW$ for each final state.
The combination of the results at $\sqrt{s}=189-209~\GeV$ yields 
for the semi-leptonic and the fully-hadronic final states: 
\begin{eqnarray}
  \label{equ:mass-qqln}
  \MW(\QQLN) & = & 80.196 \pm 0.070 \pm 0.026~\GeV \quad \mbox{and} \\ 
  \label{equ:mass-qqqq}
  \MW(\QQQQ) & = & 80.298 \pm 0.064 \pm 0.049~\GeV \quad .
\end{eqnarray}
Here and in the following, the first uncertainty is statistical and the 
second systematic.
The $\QQLN$ and the $\QQQQ$ channels exhibit a correlation of 9\%.
The contributions of the individual sources of systematic uncertainty
to the combined $\MW$ value in the $\QQLN$ channel is given in 
Table~\ref{tab:mw-syst}.

The difference between the values of $\MW$ determined in the $\QQLN$ and 
$\QQQQ$ channels is
\begin{eqnarray}
\MW({\QQLN})-\MW(\QQQQ) & = & - 0.088 \pm 0.094 \pm 0.031~\GeV \; .
\end{eqnarray}
BEC and CR effects are not included in the systematic uncertainty on the 
mass difference.
Moreover, hadronisation uncertainties are treated as uncorrelated 
between the $\QQLN$ and $\QQQQ$ final states. 
This causes the mass difference not to equal the difference of 
the mass values given in Equations (\ref{equ:mass-qqln}) 
and (\ref{equ:mass-qqqq}).

Averaging the values of the $\QQLN$ and $\QQQQ$ channels, including 
BEC and CR uncertainties and all correlations, yields
\begin{eqnarray}
  \MW(\FFFF) & = & 80.242 \pm 0.048 \pm 0.031~\GeV \; .
\end{eqnarray}
In this combination the value of $\chi^2$/d.o.f. is 29.2/27
and the weight of the fully-hadronic channel is 46\%. 
In absence of any systematic uncertainties, the statistical precision
of the measurement would be $47~\MeV$.
In Table~\ref{tab:mw-syst} the contributions of the individual sources 
of systematic uncertainty to this combined $\MW$ result are given.

The results in this paper are  combined with the direct measurements
obtained at $\sqrt{s} = 172 - 184~\GeV$~\cite{l3-direct-mw} to give
\begin{eqnarray}
  \MW(\QQLN) & = & 80.212 \pm 0.066 \pm 0.027~\GeV \quad \mbox{and} \\ 
  \MW(\QQQQ) & = & 80.325 \pm 0.061 \pm 0.052~\GeV \quad ,
\end{eqnarray}
with a correlation of 10\%.
Combining the results from direct measurements at $\sqrt{s} = 172 - 209~\GeV$
with those
result obtained from cross section measurements at 
$\sqrt{s} = 161 - 172~\GeV$~\cite{l3-threshold-mw} yields
\begin{eqnarray}
  \MW        & = & 80.270 \pm 0.046 \pm 0.031~\GeV \; .
\end{eqnarray}

The W-boson width is determined in fits for both $\MW$ and $\GW$.
Table~\ref{tab:gw-results} gives the results for $\sqrt{s}=189-209~\GeV$.
Combining all data yields
\begin{eqnarray}
  \GW        & = & 2.18 \pm 0.11 \pm 0.09~\GeV \; .
\end{eqnarray}

\clearpage

%
%

\bibliographystyle{paper303}

%
%
\newpage
\input namelist303.tex

\clearpage

\begin{table}
\begin{center}
\renewcommand{\arraystretch}{1.2}
\begin{tabular}{|c|c||c|c||c|c||c|c||c|c|}
\hline
 & &
\multicolumn{2}{|c||}{$\QQEN$} & \multicolumn{2}{|c||}{$\QQMN$} &
\multicolumn{2}{|c||}{$\QQTN$} & \multicolumn{2}{|c|}{$\QQQQ$}   \\
$\langle\sqrt{s}\rangle~[\GeV]$ & ${\cal L}~[\pb]$ &
 $N_\mathrm{data}$ & $N_\BG$ &
 $N_\mathrm{data}$ & $N_\BG$ &
 $N_\mathrm{data}$ & $N_\BG$ &
 $N_\mathrm{data}$ & $N_\BG$         \\
\hline\hline
  $   188.6   $ & $           176.8$      
& $\pz     347$ & $   \pz      22.9$      
& $\pz     341$ & $   \pz      14.9$      
& $\pz     413$ & $   \pz      69.7$      
& $       1477$ & $\pz        328.7$ \\   
  $   191.6   $ & $   \pz      29.8$      
& $\pz\pz   73$ & $   \pz\pz    4.1$      
& $\pz\pz   63$ & $   \pz\pz    2.4$      
& $\pz\pz   57$ & $   \pz      11.9$      
& $\pz     236$ & $\pz\pz      57.5$ \\   
  $   195.5   $ & $   \pz      84.1$      
& $\pz     168$ & $   \pz      10.9$      
& $\pz     157$ & $   \pz\pz    8.2$      
& $\pz     222$ & $   \pz      33.8$      
& $\pz     665$ & $\pz        153.5$ \\   
  $   199.6   $ & $   \pz      83.3$      
& $\pz     152$ & $   \pz      11.4$      
& $\pz     142$ & $   \pz\pz    7.3$      
& $\pz     181$ & $   \pz      32.2$      
& $\pz     726$ & $\pz        151.1$ \\   
  $   201.8   $ & $   \pz      37.1$      
& $\pz\pz   70$ & $   \pz\pz    5.3$      
& $\pz\pz   79$ & $   \pz\pz    3.4$      
& $\pz\pz   77$ & $   \pz      13.9$      
& $\pz     301$ & $\pz\pz      64.6$ \\   
  $   204.8   $ & $   \pz      79.0$      
& $\pz     176$ & $   \pz      11.0$      
& $\pz     142$ & $   \pz\pz    6.5$      
& $\pz     164$ & $   \pz      26.4$      
& $\pz     656$ & $\pz        137.2$ \\   
  $   206.6   $ & $           139.1$      
& $\pz     283$ & $   \pz      18.0$      
& $\pz     263$ & $   \pz      12.5$      
& $\pz     304$ & $   \pz      48.0$      
& $       1173$ & $\pz        234.2$ \\   
\hline
 Total          & $           629.4$      
& $      1269$  & $   \pz      83.6$      
& $      1187$  & $   \pz      55.2$      
& $      1418$  & $           235.9$      
& $      5234$  & $          1126.8$ \\   
\hline
\end{tabular}
\caption{
  Integrated luminosity, ${\cal L}$, together with the number of selected data events, $N_\mathrm{data}$, 
  and expected number of background events, $N_\BG$,
  for each final state and average value of $\sqrt{s}$.
  }
\label{tab:event-sample}
\end{center}
\end{table}

\begin{table}
\begin{center}
\renewcommand{\arraystretch}{1.2}
\begin{tabular}{|l||c|c|c|}
\hline
               & \makebox[2cm]{Energy~[\%]} 
               & \makebox[2cm]{$\theta$~[deg.]} 
               & \makebox[2cm]{$\phi$~[deg.]}     \\
\hline
\hline
Electrons                    & $1.4$ & $0.47$ & $0.083$ \\
Muons                        & $5.2$ & $0.22$ & $0.007$ \\
Hadronic jets (no cut)       & $15$  & $2.4 $ & $1.9  $ \\
Hadronic jets ($E_C>2~\GeV$) & $15$  & $2.5 $ & $2.1  $ \\
\hline
\end{tabular}
\caption{ 
  Average energy and angle resolutions for reconstructed electrons, 
  muons and hadronic jets as determined in Monte Carlo simulation.  
  Resolutions for hadronic jets are given with and without the cut on
  the minimum cluster energy, $E_C$.
  }
\label{tab:object-resolution}
\end{center}
\end{table}

\begin{table}
\begin{center}
\renewcommand{\arraystretch}{1.2}
\begin{tabular}{|c||c|c|c|c|c||c|}
\hline
              & $\QQEN$ & $\QQMN$ & $\QQTN$ & $\QQQQ$ & $\QQQQ$ & used in \\
Mass variable & & &                         & 4-jet   & 5-jet   & fit as  \\
\hline
\hline
$m_\mathrm{raw}^\mathrm{qq}$    
& $8.4$ & $8.5$ & $   10.8$ & $   11.6$ & $   12.4$ & \\
$m_\mathrm{raw}$ 
& $5.1$ & $7.5$ & $     - $ & $\pz 6.6$ & $\pz 6.7$ & \\
\hline
\hline
$m_\mathrm{resc}$              
& $ - $ & $ - $ & $\pz 4.4$ & $     - $ & $     - $ & $m_1$ \\
\hline
 $\Minvone$                      
& $4.7$ & $6.5$ & $     - $ & $     - $ & $     - $ & $m_2$ \\
 $\Minvtwo$                      
& $2.3$ & $2.8$ & $     - $ & $     - $ & $     - $ & $m_1$ \\
\hline
 $\Minvfour$                     
& $ - $ & $ - $ & $     - $ & $\pz 2.2$ & $\pz 3.0$ & $m_2$ \\
 $\Minvfive$                     
& $ - $ & $ - $ & $     - $ & $\pz 1.9$ & $\pz 2.5$ & $m_1$ \\
\hline
\end{tabular}
\caption{ 
  Mass resolutions in $\GeV$ as determined in Monte Carlo simulation:
  raw mass resolution, $m_\mathrm{raw}^\mathrm{qq}$, of the 
  hadronically-decaying W bosons; resolution of the average of the
  two raw masses in each event, $m_\mathrm{raw}$;
  resolution after rescaling the jet energies, $m_\mathrm{resc}$, 
  or after applying a kinematic fit, $\Minvn$. 
  The last column indicates which of the mass variables is 
  used for each final state in the extraction of $\MW$ and $\GW$,
  as described in Section~\ref{sec:fit}.
  }
\label{tab:mass-resolution}
\end{center}
\end{table}

\clearpage

\begin{table}
\begin{center}
\renewcommand{\arraystretch}{1.2}
\begin{tabular}{|l||c|c|c|c||c||c|}
\hline
                           & $\QQEN$ & $\QQMN$ & $\QQTN$ & $\QQQQ$
                                               & $\QQLN$ & $\FFFF$ \\
\hline
\hline    
Calibration of $\sqrt{s}$  &  \multicolumn{4}{|c||}{$10$}         
                           & $   10$ & $   10$                     \\
\hline    
Lepton measurement         & $\pz 6$ & $   12$ & $   - $ & $   - $ 
                                               & $\pz 5$ & $\pz 3$ \\ 
\hline    
Jet measurement            & $\pz 4$ & $   11$ & $   23$ & $\pz 5$
                                               & $\pz 9$ & $\pz 7$ \\ 
\hline               
Background                 & $\pz 2$ & $\pz 1$ & $   23$ & $\pz 7$
                                               & $\pz 3$ & $\pz 4$ \\ 
\hline    
MC statistics              & $\pz 7$ & $\pz 9$ & $   22$ & $   10$
                                               & $\pz 5$ & $\pz 6$ \\ 
\hline 
Photon radiation           & $   16$ & $   10$ & $\pz 9$ & $\pz 6$
                                               & $   13$ & $   10$ \\ 
\hline    
Hadronisation              & $   11$ & $   12$ & $   44$ & $   20$
                                               & $   16$ & $   18$ \\ 
\hline    
Bose-Einstein correlations & $   - $ & $   - $ & $   - $ & $   17$
                                               & $   - $ & $\pz 8$ \\ 
\hline    
Colour reconnection        & $   - $ & $   - $ & $   - $ & $   38$
                                               & $   - $ & $   17$ \\ 
\hline    
\hline                                              
Total systematic           & $   24$ & $   26$ & $   60$ & $   49$
                                               & $   26$ & $   31$ \\
\hline
Total statistical          & $   99$ & $119\pz$& $175\pz$& $   64$
                                               & $   70$ & $   48$ \\
\hline
\hline
Total                      & $102\pz$& $121\pz$& $185\pz$& $   81$
                                               & $   74$ & $   57$ \\
\hline
\end{tabular}
\caption{
  Systematic uncertainties on $\MW$, in $\MeV$, for the various final states.
The values refer to the complete data set at $\sqrt{s}=189-209~\GeV$
  and take into account correlations between energy points and final states.
}
\label{tab:mw-syst}
\end{center}
\end{table}

\begin{table}
\begin{center}
\renewcommand{\arraystretch}{1.2}
\begin{tabular}{|l||c|c|c|c||c||c|}
\hline
                           & $\QQEN$  & $\QQMN$  & $\QQTN$  & $\QQQQ$
                                                 & $\QQLN$  & $\FFFF$  \\
\hline
\hline    
Calibration of $\sqrt{s}$  &  \multicolumn{4}{|c||}{$<\!5$}            
                           & $  <\!5$ & $  <\!5$                       \\
\hline    
Lepton measurement         & $   10 $ & $   35 $ & $   -  $ & $   -  $  
                                                 & $   15 $ & $\pz 5 $ \\ 
\hline    
Jet measurement            & $   20 $ & $   30 $ & $   75 $ & $   20 $  
                                                 & $   30 $ & $   25 $ \\ 
\hline    
Background                 & $   20 $ & $\pz 5 $ & $   45 $ & $   50 $  
                                                 & $   10 $ & $   25 $ \\ 
\hline    
MC statistics              & $   15 $ & $   20 $ & $   50 $ & $   15 $  
                                                 & $   15 $ & $   10 $ \\ 
\hline 
Photon radiation           & $\pz 5 $ & $\pz 5 $ & $\pz 5 $ & $\pz 5 $  
                                                 & $\pz 5 $ & $\pz 5 $ \\ 
\hline    
Hadronisation              & $   55 $ & $   70 $ & $150\pz$ & $   85 $  
                                                 & $   75 $ & $   80 $ \\ 
\hline    
Bose-Einstein correlations & $   -  $ & $   -  $ & $   -  $ & $   10 $  
                                                 & $   -  $ & $\pz 5 $ \\ 
\hline    
Colour reconnection        & $   -  $ & $   -  $ & $   -  $ & $   50 $  
                                                 & $   -  $ & $   25 $ \\ 
\hline    
\hline                                              
Total systematic           & $  65  $ & $   90 $ & $180\pz$ & $115\pz$ 
                                                 & $   85 $ & $   90 $ \\
\hline
Total statistical          & $245\pz$ & $305\pz$ & $380\pz$ & $150\pz$ 
                                                 & $170\pz$ & $115\pz$ \\
\hline
\hline
Total                      & $255\pz$ & $315\pz$ & $420\pz$ & $190\pz$ 
                                                 & $190\pz$ & $145\pz$ \\
\hline
\end{tabular}
\caption{
  Systematic uncertainties on $\GW$, in $\MeV$, for the various final states.
  All uncertainties are rounded to the next $5~\MeV$.
The values refer to the complete data set at $\sqrt{s}=189-209~\GeV$
  and take into account correlations between energy points and final states.
}
\label{tab:gw-syst}
\end{center}
\end{table}

\clearpage

\begin{table}
\begin{center}
\renewcommand{\arraystretch}{1.2}
\begin{tabular}{|l||c|c|c|c||c|c|c|c|}
\hline
              & \multicolumn{4}{|c||}{$|\Delta\MW|~[\MeV]$}
              & \multicolumn{4}{|c|}{$|\Delta\GW|~[\MeV]$} \\
\cline{2-9}
              & $\QQEN$ & $\QQMN$ & $\QQTN$ & $\QQQQ$ 
              & $\QQEN$ & $\QQMN$ & $\QQTN$ & $\QQQQ$    \\
\hline
\hline
Electron energy
              & $\pz 6$ & $   - $ & $   - $ & $   - $     
              & $   12$ & $   - $ & $   - $ & $   - $  \\ 
Muon energy
              & $   - $ & $   12$ & $   - $ & $   - $     
              & $   - $ & $   37$ & $   - $ & $   - $  \\ 
Jet energy scale ($\pm 50~\MeV$) 
              & $\pz 3$ & $   10$ & $\pz 9$ & $\pz 2$     
              & $\pz 1$ & $\pz 9$ & $   16$ & $\pz 4$  \\ 
Jet energy smearing (1\%)         
              & $\pz 1$ & $\pz 4$ & $\pz 7$ & $\pz 1$     
              & $   10$ & $   25$ & $   53$ & $\pz 7$  \\ 
Jet angle smearing ($0.5^\circ$)         
              & $\pz 2$ & $\pz 4$ & $   20$ & $\pz 4$     
              & $   17$ & $   16$ & $   47$ & $   18$  \\ 
\hline
\end{tabular}
\caption{
  Changes of $\MW$ and $\GW$ due to variations of the energy 
  measurement of electrons, muons and jets
  and the resolutions of the jet directions.
  }
\label{tab:detector}
\end{center}
\end{table}

\begin{table}
\begin{center}
\renewcommand{\arraystretch}{1.2}
\begin{tabular}{|l||c|c|c|c||c|c|c|c|}
\hline
              & \multicolumn{4}{|c||}{$|\Delta\MW|~[\MeV]$}
              & \multicolumn{4}{|c|}{$|\Delta\GW|~[\MeV]$} \\
\cline{2-9}
              & $\QQEN$ & $\QQMN$ & $\QQTN$ & $\QQQQ$ 
              & $\QQEN$ & $\QQMN$ & $\QQTN$ & $\QQQQ$    \\
\hline
\hline
Four-fermion background   
              & $\pz 2$    & $\pz 1$ & $\pz 2$ & $\pz 3$   
              & $   12$    & $\pz 1$ & $   12$ & $   11$ \\
$\EE\to\QQ$ background         
              & $<\!\!1\,$ & $<\!\!1\,$& $ 23$ & $\pz 3$   
              & $   15$    & $\pz 6$ & $   44$ & $   31$ \\
$y_{34}$ spectrum
              & $   - $    & $   - $ & $   - $ & $\pz 5$   
              & $   - $    & $   - $ & $   - $ & $   40$ \\
\hline
\end{tabular}
\caption{
  Changes of $\MW$ and $\GW$ due to variations of the background processes.
  For fully-hadronic events the uncertainty due to the $y_{34}$ spectrum 
  is also given.
  }
\label{tab:background}
\end{center}
\end{table}

\begin{table}
\begin{center}
\renewcommand{\arraystretch}{1.2}
\begin{tabular}{|l||c|c|c|c||c|c|c|c|}
\hline
                   & \multicolumn{4}{|c||}{$\Delta\MW~[\MeV]$}
                   & \multicolumn{4}{|c|}{$\Delta\GW~[\MeV]$} \\
\cline{2-9}
                   & $\QQEN$ & $\QQMN$ & $\QQTN$ & $\QQQQ$ 
                   & $\QQEN$ & $\QQMN$ & $\QQTN$ & $\QQQQ$  \\
\hline
\hline
Generator comparison     
           & $   {-16}$ & $\pz {+9}$ & $     -  $ & $\pz {+4}$     
           & $     -  $ & $     -  $ & $     -  $ & $     -  $  \\ 
Monte Carlo reweighting 
           & $\pz\phantom{+}{0}$ 
                        & $\pz {-5}$ & $\pz {-1}$ & $\pz {+5}$     
           & $\pz {+3}$ & $\pz {+5}$ & $\pz {-3}$ & $\pz {-3}$  \\ 
\hline
\end{tabular}
\caption{
  Changes of $\MW$ and $\GW$ due to variations in the modelling of 
  photon radiation.
  The first row gives the results of a comparison between the 
  RACOONWW and KANDY generators.
  The second row gives the difference between the ${\cal O}(\alpha^3)$
  and the ${\cal O}(\alpha^2)$ calculation, obtained by 
  reweighting KANDY events.
  The statistical accuracy of the generator comparison is about $8~\MeV$,
  while the statistical uncertainty of the reweighting procedure is negligible.
}
\label{tab:photons}
\end{center}
\end{table}

\clearpage

\begin{table}
\begin{center}
\renewcommand{\arraystretch}{1.2}
\begin{tabular}{|l||c|c|c|c||c|c|c|c|}
\hline
              & \multicolumn{4}{|c||}{$\Delta\MW~[\MeV]$}
              & \multicolumn{4}{|c|}{$\Delta\GW~[\MeV]$} \\
\cline{2-9}
         & $\QQEN$ & $\QQMN$ & $\QQTN$ & $\QQQQ$
         & $\QQEN$ & $\QQMN$ & $\QQTN$ & $\QQQQ$ \\
\hline
\hline
HERWIG   & $\pz\phantom{+}{0}$ 
                        & $\pz  {-8}$ & $    {-41}$ & $\pz  {-3}$    
         & $\pz  {-96}$ & $   {-141}$ & $   {-275}$ & $\pz {-32}$ \\ 
ARIADNE  & $     {-15}$ & $    {-11}$ & $    {-44}$ & $    {+11}$    
         & $\pz  {+15}$ & $\pzz {-1}$ & $\pz {-24}$ & $\pzz {+5}$ \\ 
\hline
\end{tabular}
\caption{
  Changes of $\MW$ and $\GW$ due to the use of the hadronisation models
  HERWIG and ARIADNE instead of PYTHIA.
  The statistical accuracy is always better than 
  $15~\MeV$ and $30~\MeV$ for $\MW$ and $\GW$, respectively.
  }
\label{tab:hadronisation}
\end{center}
\end{table}

\begin{table}
\begin{center}
\renewcommand{\arraystretch}{1.2}
\begin{tabular}{|l||c|c||c|c|}
\hline
              & \multicolumn{2}{|c||}{$\Delta\MW~[\MeV]$}
              & \multicolumn{2}{|c|}{$\Delta\GW~[\MeV]$} \\
\cline{2-5}
              & $\QQLN$ & $\QQQQ$ 
              & $\QQLN$ & $\QQQQ$    \\
\hline
\hline
Kaon multiplicity   & $   {+13}\pm   12$ & $   {+25}\pm 23$      
                    & $   {-12}\pm   11$ & $   {-95}\pm 87$   \\ 
Proton multiplicity & $\pz {+1}\pm\pz 2$ & $\pz {+7}\pm 15$      
                    & $\pz {-3}\pm\pz 5$ & $   {-36}\pm 80$   \\ 
\hline
\end{tabular}
\caption{
  Changes of $\MW$ and $\GW$ due to reweighting Monte Carlo events with 
  respect to variations of the mean charged-kaon and proton multiplicities. 
  The given uncertainties are due to the experimental uncertainties in the
  determination of these multiplicities~\cite{delphi-kaon-proton}.
  }
\label{tab:reweighting}
\end{center}
\end{table}

\begin{table}
\begin{center}
\renewcommand{\arraystretch}{1.2}
\begin{tabular}{|c||c||c|}
\hline
$\sqrt{s}~[\GeV]$  & $\Delta\MW~[\MeV]$ & $\Delta\GW~[\MeV]$  \\
\hline
\hline
189                & $   {+11}$          
                   & $\pz {-1}$       \\ 
207                & $   {+23}$          
                   & $   {-15}$       \\ 
\hline
\end{tabular}
\caption{
  Changes of $\MW$ and $\GW$ in the $\QQQQ$ channel when replacing our
  standard simulation by the PYTHIA BE32 model with a strength of 
  inter-W BEC corresponding to the 68\% upper limit set by our
  direct BEC measurement~\cite{l3-257-BE}.
  The statistical accuracy is $6~\MeV$ and $14~\MeV$ for
  $\MW$ and $\GW$, respectively.
  }
\label{tab:BE}
\end{center}
\end{table}

\begin{table}
\begin{center}
\renewcommand{\arraystretch}{1.2}
\begin{tabular}{|c||c||c|}
\hline
$\sqrt{s}~[\GeV]$  & $\Delta\MW~[\MeV]$ & $\Delta\GW~[\MeV]$  \\
\hline
\hline
189                & $   {-22}$          
                   & $   {-48}$       \\ 
207                & $   {-57}$          
                   & $   {-56}$       \\ 
\hline
\end{tabular}
\caption{
  Changes of $\MW$ and $\GW$ in the $\QQQQ$ final state when replacing our
  standard simulation by the PYTHIA SK-I model with $k_\mathrm{I}=1.1$ 
  which is the 68\% upper limit set by our CR measurement~\cite{l3-266-CR}.
  The cut on the minimum cluster energy of $2~\GeV$ is applied.
  The statistical accuracies are about $10~\MeV$ for $\MW$ and 
  $20~\MeV$ for $\GW$.
  }
\label{tab:CR}
\end{center}
\end{table}

\clearpage

\begin{table}
\begin{center}
\renewcommand{\arraystretch}{1.2}
\begin{tabular}{|l||c||c|}
\hline
\multicolumn{1}{|c||}{Process}  &  $\MW~[\GeV]$                  
               & $\sigma^\mathrm{exp}_\mathrm{stat}~[\GeV]$   \\
\hline\hline
$\EEQQEN$  &       $80.225 \pm 0.099 \pm 0.024$  & $0.095$ \\
$\EEQQMN$  &       $80.152 \pm 0.119 \pm 0.026$  & $0.119$ \\
$\EEQQTN$  &       $80.195 \pm 0.175 \pm 0.060$  & $0.162$ \\
\hline                               
$\EEQQLN$  &       $80.196 \pm 0.070 \pm 0.026$  & $0.068$ \\
\hline                               
$\EEQQQQ$  &       $80.298 \pm 0.064 \pm 0.049$  & $0.062$ \\
\hline                                
$\EEFFFF$  &       $80.242 \pm 0.048 \pm 0.031$  & $0.047$ \\
\hline
\end{tabular}
\caption{
  Results on $\MW$ for the data collected at $\sqrt{s} = 189-209~\GeV$.  
  The first uncertainty is statistical and the second systematic.  
  Also shown is the expected statistical uncertainty, 
  $\sigma^\mathrm{exp}_\mathrm{stat}$.
  }
\label{tab:mw-results}
\end{center}
\end{table}

\begin{table}
\begin{center}
\renewcommand{\arraystretch}{1.2}
\begin{tabular}{|l||c|c||c|}
\hline
\multicolumn{1}{|c||}{Process}  &  $\MW~[\GeV]$        
               & $\GW~[\GeV]$   &  Correlation \\
\hline\hline
$\EEQQLN$ & $80.174 \pm 0.078 \pm0.027$ & $2.50 \pm 0.17 \pm 0.09$ & $0.01$ \\
$\EEQQQQ$ & $80.284 \pm 0.074 \pm0.050$ & $1.97 \pm 0.15 \pm 0.12$ & $0.15$ \\
\hline                                              
$\EEFFFF$ & $80.236 \pm 0.054 \pm0.032$ & $2.22 \pm 0.11 \pm 0.09$ & $0.14$ \\
 \hline
\end{tabular}
\caption{
  Results on $\MW$ and $\GW$ obtained from a simultaneous fit of both
  quantities using data collected at $\sqrt{s}=189-209~\GeV$. 
  The first uncertainty is statistical and the second systematic.
  Also quoted is the correlation between $\MW$ and $\GW$.
  }
\label{tab:gw-results}
\end{center}
\end{table}

\clearpage

\clearpage

\clearpage

\clearpage

\clearpage

\clearpage


\clearpage

\clearpage

\clearpage

\clearpage

\clearpage

\clearpage

\end{document}

%% file: namelist303.tex
\typeout{   }     
\typeout{Using author list for paper 287 -  }
\typeout{$Modified: Jul 15 2001 by smele $}
\typeout{!!!!  This should only be used with document option a4p!!!!}
\typeout{   }
%
%
%
%
%
%

\newcount\tutecount  \tutecount=0
\def\tutenum#1{\global\advance\tutecount by 1 \xdef#1{\the\tutecount}}
\def\tute#1{$^{#1}$}
\tutenum\aachen            
\tutenum\nikhef            
\tutenum\mich              
\tutenum\lapp              
\tutenum\basel             
\tutenum\lsu               
\tutenum\beijing           
\tutenum\bologna           
\tutenum\tata              
\tutenum\ne                
\tutenum\bucharest         
\tutenum\budapest          
\tutenum\mit               
\tutenum\panjab            
\tutenum\debrecen          
\tutenum\dublin            
\tutenum\florence          
\tutenum\cern              
\tutenum\wl                
\tutenum\geneva            
\tutenum\hamburg           
\tutenum\hefei             
\tutenum\lausanne          
\tutenum\lyon              
\tutenum\madrid            
\tutenum\florida           
\tutenum\milan             
\tutenum\moscow            
\tutenum\naples            
\tutenum\cyprus            
\tutenum\nymegen           
\tutenum\caltech           
\tutenum\perugia           
\tutenum\peters            
\tutenum\cmu               
\tutenum\potenza           
\tutenum\prince            
\tutenum\riverside         
\tutenum\rome              
\tutenum\salerno           
\tutenum\ucsd              
\tutenum\sofia             
\tutenum\korea             
\tutenum\taiwan            
\tutenum\tsinghua          
\tutenum\purdue            
\tutenum\psinst            
\tutenum\zeuthen           
\tutenum\eth               

{
\parskip=0pt
\noindent
{\bf The L3 Collaboration:}
\ifx\selectfont\undefined
 \baselineskip=10.8pt
 \baselineskip\baselinestretch\baselineskip
 \normalbaselineskip\baselineskip
 \ixpt
\else
 \fontsize{9}{10.8pt}\selectfont
\fi
\medskip
\tolerance=10000
\hbadness=5000
\raggedright
\hsize=162truemm\hoffset=0mm
\def\r{\rlap,}
\noindent

P.Achard\r\tute\geneva\ 
O.Adriani\r\tute{\florence}\ 
M.Aguilar-Benitez\r\tute\madrid\ 
J.Alcaraz\r\tute{\madrid}\ 
G.Alemanni\r\tute\lausanne\
J.Allaby\r\tute\cern\
A.Aloisio\r\tute\naples\ 
M.G.Alviggi\r\tute\naples\
H.Anderhub\r\tute\eth\ 
V.P.Andreev\r\tute{\lsu,\peters}\
% ...
\newpage
\begin{list}{A}{\itemsep=0pt plus 0pt minus 0pt\parsep=0pt plus 0pt minus 0pt
                \topsep=0pt plus 0pt minus 0pt}
\item[\aachen]
 III. Physikalisches Institut, RWTH, D-52056 Aachen, Germany$^{\S}$
\item[\nikhef] National Institute for High Energy Physics, NIKHEF, 
     and University of Amsterdam, NL-1009 DB Amsterdam, The Netherlands
\item[\mich] University of Michigan, Ann Arbor, MI 48109, USA
\item[\lapp] Laboratoire d'Annecy-le-Vieux de Physique des Particules, 
     LAPP,IN2P3-CNRS, BP 110, F-74941 Annecy-le-Vieux CEDEX, France
\item[\basel] Institute of Physics, University of Basel, CH-4056 Basel,
     Switzerland
\item[\lsu] Louisiana State University, Baton Rouge, LA 70803, USA
\item[\beijing] Institute of High Energy Physics, IHEP, 
  100039 Beijing, China$^{\triangle}$ 
\item[\bologna] University of Bologna and INFN-Sezione di Bologna, 
     I-40126 Bologna, Italy
\item[\tata] Tata Institute of Fundamental Research, Mumbai (Bombay) 400 005, India
\item[\ne] Northeastern University, Boston, MA 02115, USA
\item[\bucharest] Institute of Atomic Physics and University of Bucharest,
     R-76900 Bucharest, Romania
\item[\budapest] Central Research Institute for Physics of the 
     Hungarian Academy of Sciences, H-1525 Budapest 114, Hungary$^{\ddag}$
\item[\mit] Massachusetts Institute of Technology, Cambridge, MA 02139, USA
\item[\panjab] Panjab University, Chandigarh 160 014, India
\item[\debrecen] KLTE-ATOMKI, H-4010 Debrecen, Hungary$^\P$
\item[\dublin] UCD School of Physics, University College Dublin, 
 Belfield, Dublin 4, Ireland
\item[\florence] INFN Sezione di Firenze and University of Florence, 
     I-50125 Florence, Italy
\item[\cern] European Laboratory for Particle Physics, CERN, 
     CH-1211 Geneva 23, Switzerland
\item[\wl] World Laboratory, FBLJA  Project, CH-1211 Geneva 23, Switzerland
\item[\geneva] University of Geneva, CH-1211 Geneva 4, Switzerland
\item[\hamburg] University of Hamburg, D-22761 Hamburg, Germany
\item[\hefei] Chinese University of Science and Technology, USTC,
      Hefei, Anhui 230 029, China$^{\triangle}$
\item[\lausanne] University of Lausanne, CH-1015 Lausanne, Switzerland
\item[\lyon] Institut de Physique Nucl\'eaire de Lyon, 
     IN2P3-CNRS,Universit\'e Claude Bernard, 
     F-69622 Villeurbanne, France
\item[\madrid] Centro de Investigaciones Energ{\'e}ticas, 
     Medioambientales y Tecnol\'ogicas, CIEMAT, E-28040 Madrid,
     Spain${\flat}$ 
\item[\florida] Florida Institute of Technology, Melbourne, FL 32901, USA
\item[\milan] INFN-Sezione di Milano, I-20133 Milan, Italy
\item[\moscow] Institute of Theoretical and Experimental Physics, ITEP, 
     Moscow, Russia
\item[\naples] INFN-Sezione di Napoli and University of Naples, 
     I-80125 Naples, Italy
\item[\cyprus] Department of Physics, University of Cyprus,
     Nicosia, Cyprus
\item[\nymegen] Radboud University and NIKHEF, 
     NL-6525 ED Nijmegen, The Netherlands
\item[\caltech] California Institute of Technology, Pasadena, CA 91125, USA
\item[\perugia] INFN-Sezione di Perugia and Universit\`a Degli 
     Studi di Perugia, I-06100 Perugia, Italy   
\item[\peters] Nuclear Physics Institute, St. Petersburg, Russia
\item[\cmu] Carnegie Mellon University, Pittsburgh, PA 15213, USA
\item[\potenza] INFN-Sezione di Napoli and University of Potenza, 
     I-85100 Potenza, Italy
\item[\prince] Princeton University, Princeton, NJ 08544, USA
\item[\riverside] University of Californa, Riverside, CA 92521, USA
\item[\rome] INFN-Sezione di Roma and University of Rome, ``La Sapienza",
     I-00185 Rome, Italy
\item[\salerno] University and INFN, Salerno, I-84100 Salerno, Italy
\item[\ucsd] University of California, San Diego, CA 92093, USA
\item[\sofia] Bulgarian Academy of Sciences, Central Lab.~of 
     Mechatronics and Instrumentation, BU-1113 Sofia, Bulgaria
\item[\korea]  The Center for High Energy Physics, 
     Kyungpook National University, 702-701 Taegu, Republic of Korea
\item[\taiwan] National Central University, Chung-Li, Taiwan, China
\item[\tsinghua] Department of Physics, National Tsing Hua University,
      Taiwan, China
\item[\purdue] Purdue University, West Lafayette, IN 47907, USA
\item[\psinst] Paul Scherrer Institut, PSI, CH-5232 Villigen, Switzerland
\item[\zeuthen] DESY, D-15738 Zeuthen, Germany
\item[\eth] Eidgen\"ossische Technische Hochschule, ETH Z\"urich,
     CH-8093 Z\"urich, Switzerland
\item[\S]  Supported by the German Bundesministerium 
        f\"ur Bildung, Wissenschaft, Forschung und Technologie.
\item[\ddag] Supported by the Hungarian OTKA fund under contract
numbers T019181, F023259 and T037350.
\item[\P] Also supported by the Hungarian OTKA fund under contract
  number T026178.
\item[$\flat$] Supported also by the Comisi\'on Interministerial de Ciencia y 
        Tecnolog{\'\i}a.
\item[$\sharp$] Also supported by CONICET and Universidad Nacional de La Plata,
        CC 67, 1900 La Plata, Argentina.
\item[$\triangle$] Supported by the National Natural Science
  Foundation of China.
\end{list}
}
\vfill
